﻿
\documentclass[aip,jcp,reprint,amssymb,amsmath,superscriptaddress,groupedaddress,frontmatterverbose,]{revtex4-2}

\usepackage{amsmath}
\usepackage{amssymb}
\usepackage{bm}
\usepackage{color}
\usepackage{braket}
\usepackage{graphicx}
\usepackage{orcidlink}
\usepackage{epstopdf}
\usepackage{physics}
\usepackage{comment}
\DeclareGraphicsExtensions{{.eps,.pdf,.png}}
\graphicspath{{./}}

\begin{document}
\title{Simulating two-dimensional correlation spectroscopies with third-order infrared and fifth-order infrared--Raman processes of liquid water}
\date{Last updated: \today}

\author{Hideaki Takahashi\orcidlink{0000-0001-6465-2049}}
\author{Yoshitaka Tanimura\orcidlink{0000-0002-7913-054X}}
\email[Author to whom correspondence should be addressed: ]{tanimura.yoshitaka.5w@kyoto-u.jp}
\affiliation{Department of Chemistry, Graduate School of Science,
Kyoto University, Kyoto 606-8502, Japan}

\begin{abstract}
To investigate the possibility of measuring the intermolecular and intramolecular anharmonic coupling of balk water, we calculate third-order two-dimensional (2D) infrared (IR) spectra and fifth-order 2D IR--IR--Raman--Raman spectra expressed in terms of four-body correlation functions of optical observables. For this purpose, a multimode Brownian oscillator model of four interacting anharmonic oscillators strongly coupled to their respective heat baths is employed. The nonlinearity of the system--bath interactions is considered to describe thermal relaxation and vibrational dephasing. The linear and nonlinear spectra are then computed in a non-Markovian and nonperturbative regime in a rigorous manner using the discretized hierarchical equations of motion in mixed Liouville--Wigner space (DHEOM--MLWS). The calculated 2D spectra for stretching--bending, bending--librational, stretching--librational, and stretching--translational modes consist of various positive and negative peaks exhibiting essential details of the intermolecular and intramolecular mode--mode interactions under thermal relaxation and dephasing at finite temperature.\end{abstract}
\maketitle

\section{INTRODUCTION}
The delineation of the spectroscopic line shapes of liquid water into vibrational modes and their interactions has been the subject of numerous experimental and theoretical studies.\cite{OCSACR1999,Nibbering2004UltrafastVD,SkinnerCPL2004, BakkerSkinner2010} The uniqueness of water arises from its high-frequency intramolecular modes that promote bond forming and bond breaking through complex hydrogen bonding (HB)\cite{Tokmakoff2003H2O, Hynes2004,Imoto_JCP135,NagataChemRev2016} and its low-frequency intermolecular modes that realize irreversible nuclear motion.\cite{Ohmine_ChemRev93,Yagasaki_ACR42,Yagasaki_ARPC64,Imotobend-lib2015} The interplay between these modes plays key roles in chemical reaction processes.\cite{ElsaesserCPL2005,TokmakoffNat2013} 

Experimentally, infrared (IR)\cite{Bertie96IRexp,IRexp2011} and third-order off-resonant Raman spectroscopies\cite{Brooker1989Ramanexp,Raman2018exp, HEISLER201745} have been utilized for the analysis of the vibrational motions of liquid water. The observables of these methods are defined in terms of the dipole $\hat{\bm{\mu}}$ and polarizability $\hat{\bm{\Pi}}$ as $R_{\mathrm{IR}}^{(1)}(t_1)=i\langle [\hat{\bm{\mu}} (t_{1}), \hat{\bm{\mu }} ]\rangle /\hbar$ and $R_{\mathrm{Raman}}^{(3)}(t_1)=i\langle [\hat{\bm{\Pi }}(t_{1}), \hat{\bm{\Pi }}]\rangle /\hbar$, respectively, as functions of a single time duration $t_1$, which is referred to as one-dimensional (1D) spectroscopy.\cite{mukamel1999principles} 
The IR process has been utilized not only for investigations of the intramolecular OH stretching motion ($\sim$3700 cm$^{-1}$) and the intramolecular HOH bending motion ($\sim$1600 cm$^{-1}$), but also those of the HB intermolecular translation (vibrational) motion (33$\sim$100 cm$^{-1}$) and inter-HB molecular vibrational motion (33$\sim$400 cm$^{-1}$), which are also regarded as THz processes. The Raman process, which exploits the large polarizability of water molecules, has been used to probe both intramolecular and intermolecular modes. This approach is particularly useful for investigations of intermolecular modes because of its superior time resolution and intensity in such a low-frequency region. Although 1D spectroscopic approaches are useful for the classification of vibrational line shapes of liquids into intermolecular and intramolecular modes, it is unclear whether the peaks that we investigate are mutually coupled, and whether the width of the peaks that we measure are of an inhomogeneous origin or a homogeneous origin, as these peaks are usually broad and overlap.

Two-dimensional (2D) experiments, in which an additional time correlation is imprinted on the system response, are capable of determining anharmonic coupling among vibrational modes and homogeneous linewidths that are needed to understand the degree of coupling to the bath (i.e., the surrounding molecules with respect to the excited transition).\cite{TM93JCP,mukamel1999principles,TI09ACR,Cho2009,Hamm2011ConceptsAM} Recent developments of experimental techniques have significantly augmented the study of both intermolecular and intramolecular modes in the 0$\sim$4000 cm$^{-1}$ frequency range through 2D IR,\cite{ElsaesserDwaynePNAS2008,Tokmakoff2016H2O,VothTokmakoff_St-BendJCP2017,Tokmakoff2022,Kuroda_BendPCCP2014} 2D THz--Raman spectroscopies,\cite{HammTHz2012,Hamm2013PNAS,hamm2014,HammPerspH2O2017} and 2D THz--IR--visible (TIV) spectroscopy,\cite{grechko2018,Bonn2DTZIFvis2021} which is equivalent to 2D IR--IR--Raman spectroscopy.\cite{IT16JCP,TT23JCP1} 
Although 2D IR and 2D THz--Raman experiments provide information about intramolecular and intermolecular degrees of freedom, respectively, neither approach can directly measure the anharmonic coupling between intermolecular and intramolecular modes. In contrast, 2D TIV, which utilizes light sources to excite both intermolecular and intramolecular modes, allows us to investigate their mode--mode coupling mechanism.
Note that the analysis of both 2D IR-Raman and 2D THz--Raman spectroscopies defined in terms of three-body response functions such as $R_{\mathrm{IIR}}^{(3)}(t_{2},t_{1})=-\langle [[\hat{\bm{\Pi }}(t_{2}+t_{1}),\hat{\bm{\mu }}(t_{1})],\hat{\bm{\mu }}(0)]\rangle /\hbar ^{2}$ is far more complex than that of third-order 2D IR spectroscopy defined in terms of four-body correlation functions such as $R_{\mathrm{2DIR}}^{(3)}(t_{3},t_{2},t_{1})=-i\langle [[\hat{\bm{\mu}}(t_{3}+t_{2}+t_{1}), [[\hat{\bm{\mu }}(t_{2}+t_{1}),\hat{\bm{\mu }}(t_{1})],\hat{\bm{\mu }}(0)]]\rangle /\hbar ^{3}$ because the contributions from the nonlinear polarizability and dipole moments involved in the laser interactions give rise to additional peaks, and nondiagonal spectral peaks are not necessary to represent mode--mode coupling peaks, as in the case of 2D IR spectra.\cite{OT03JPC,Hamm2DTHzFeynmann}

In this study, we extend our analysis of 2D THz--Raman\cite{IHT14JCP,IJT15SD, IHT16JPCL,IIT15JCP} and 2D IR--Raman spectroscopies\cite{IT16JCP,TT23JCP1} by including an additional time duration for laser excitation and calculate the fifth-order 2D IR--IR--Raman--Raman spectrum defined as $R_{\mathrm{IIRR}}^{(5)}(t_{3},t_{2},t_{1})=-i\langle [[\hat{\bm{\Pi}}( t_{3}+t_{2}+t_{1}), [[\hat{\bm{\Pi }}(t_{2}+t_{1}),\hat{\bm{\mu }}(t_{1})],\hat{\bm{\mu }}(0)]]\rangle /\hbar ^{3}$, in addition to the third-order 2D IR spectrum defined as $R_{\mathrm{2DIR}}^{(3)}(t_{3},t_{2},t_{1})$. Through simulation, we attempt to clarify the roles of anharmonic mode--mode coupling and vibrational dephasing, as well as thermal relaxation, among the intermolecular and intramolecular modes of balk water. 

It should be noted that simulating the 2D spectrum of liquid water for a mixture of intramolecular and intermolecular modes is challenging, especially on account of the peak-splitting in third-order 2D IR spectra caused by the transitions between the $|0\rangle$ $\rightarrow$ $|1\rangle$ $\rightarrow$ $|2\rangle$ and $|0\rangle$ $\rightarrow$ $|1\rangle$ $\rightarrow$ $|0\rangle$ states, where $|n\rangle$ is the $n$th vibrational energy state.\cite{IT06JCP,ST11JPCA,ChoH2OMD2014} The high-frequency intramolecular motions must be addressed within the framework of quantum dynamics theory, and their dynamics involve complex hydrogen bond networks for 2D spectroscopic measurements. These motions have only been studied from classical MD simulations\cite{ImotXanteasSaitoJCP2013H2O,ChoH2OMD2014} or classical MD-based quantum simulations, such as using stochastic theory\cite{SkinnerStochs2003} and wave function–based theory.\cite{Mukamel2009,JansenSkiner2010} Therefore, the peak profiles of experimentally obtained 2D IR spectra--particularly for mode--mode coupling peaks\cite{ElsaesserDwaynePNAS2008,Tokmakoff2016H2O,VothTokmakoff_St-BendJCP2017,Tokmakoff2022}--have not been accurately predicted from a theoretical approach. 

As a practical approach, the Brownian oscillator (BO) model has been employed for the analysis of both linear\cite{Kuehn2015JPCL, Kuehn2015JCP,Kuehn2016JCP} and nonlinear spectra.\cite{TM93JCP,mukamel1999principles,TI09ACR} In this approach, the vibrational modes representing the spectroscopic properties of interest are described as functions of molecular coordinates, whereas the environmental molecular motions are described using heat baths that exert thermal fluctuations and dissipation on the vibrational modes. Although analyses based on MD simulations have shown that vibrational relaxation and dephasing are important mechanisms for characterizing molecular motions,\cite{Yagasakirelax2009,YagasakiSaitoJCP2011Relax} the inclusion of the nonlinear and non-Markovian interactions with the heat baths,\cite{OT97PRE,TS20JPSJ,KT04JCP} as well as the anharmonic mode--mode interactions,\cite{IT07JPCA,TI09ACR} are also significant. Therefore, for the vibrational motions of liquid water, we introduced the multimode nonlinear BO model, which consists of interacting anharmonic oscillators coupled to their respective heat baths with linear--linear (LL) and square--linear (SL) system--bath (SB) interactions to describe vibrational relaxation and dephasing.\cite{IIT15JCP,IT16JCP,TT23JCP1} To guarantee the equilibrium state of the system at finite temperature, the model must be solved accurately such that the dephasing (fluctuations) and relaxation (dissipation) satisfy the quantum version of the fluctuation-dissipation theorem.\cite{TK89JPSJ1,T06JPSJ,T20JCP}
 
The model parameters were then chosen to fit the classical MD simulation for 2D IR--Raman spectra conducted using the polarizable water model for inter- and intramolecular vibrational spectroscopy (POLI2VS);\cite{HT11JPCB} we then modified the anharmonicity of intramolecular modes to fit experimentally obtained 1D Raman and IR spectra,\cite{TT23JCP1} which agree with quantum MD results obtained using the POLI2VS model\cite{JianLiu2018H2OMP} and with the ab-initio MD results that have recently become available. \cite{BowmanQM_QD2015,Paesan2018H2OCMD,Althorpe2019CMD} To simulate 2D correlation spectra that detect the relaxation time of vibrational dephasing, it is necessary to deal with non-perturbative and non-Markovian SB interactions, especially for intramolecular modes, in which vibrational motion and the environment are quantum-mechanically entangled.\cite{IT06JCP,ST11JPCA} Although various hierarchical equations of motion (HEOM) approaches have been developed for investigating nonlinear spectroscopies,\cite{TK89JPSJ1,T06JPSJ,T20JCP} here we employ discretized hierarchical equations of motion in mixed Liouville--Wigner space (DHEOM--MLWS) specifically developed for the water BO model.\cite{TT23JCP1}

This paper is organized as follows.
In Sec.~\ref{sec:model}, we explain our multimode LL+SL BO model and the parameter set used for simulations. In Sec.~\ref{sub:HEOM2D}, we give the definitions of the nonlinear optical observables to be computed.
 In Sec. ~\ref{result}, we present the results for third-order 2D correlation IR and fifth-order 2D correlation IR--IR--Raman--Raman spectra based on two-mode calculations. Section~\ref{sec:conc} is devoted to our concluding remarks.

\section{Nonlinear BO model for vibrational modes of Water}\label{sec:model}
\label{sub:Model}
The details of the model and the DHEOM--MLWS for the quantum case have been explained in Refs. \onlinecite{IT16JCP,TT23JCP1}. Here, we present an overview of the model because it is helpful for the analysis of calculated 2D spectra.

\begin{table*}[!tb]
  \caption{\label{tab:FitAll1} Parameter values of multimode LL+SL BO model for (1)  stretching, (2) bending, (3) librational, and (4) translational modes obtained on the basis of Refs. \onlinecite{IT16JCP,TT23JCP1}. The normalized parameters are defined as $\tilde{\zeta}_s \equiv (\omega_0/\omega_s)^2\zeta_s$, $\tilde{V}_{LL}^{(s)} \equiv (\omega_s/\omega_0)V_{LL}^{(s)}$, $\tilde{V}_{SL}^{(s)} \equiv V_{SL}^{(s)}$, $\tilde{g}_{s^3} \equiv (\omega_s/\omega_0)^3 g_{s^3}$,$\tilde{\mu}_{s} \equiv (\omega_0/\omega_s)\mu_{s}$, $\tilde{\mu}_{ss} \equiv (\omega_0/\omega_s)^2 \mu_{ss}$, $\tilde{\Pi}_{s} \equiv (\omega_0/\omega_s)\Pi_{s}$, and $\tilde{\Pi}_{ss} \equiv (\omega_0/\omega_s)^2 \Pi_{ss}$, where we set the fundamental frequency as $\omega_{0} = 4000$~cm$^{-1}$.}
\scalebox{0.9}{
\begin{tabular}{ccccccccccc}
  \hline \hline
s&    $\omega_s$ (cm$^{-1}$) & $\gamma_s/\omega_0$ & $\tilde{\zeta}_s$ & $\tilde{V}_{LL}^{(s)}$ & $\tilde{V}_{SL}^{(s)}$ & $\tilde{g}_{s^3}$ & $\tilde{\mu}_{s}$  & $\tilde{\mu}_{ss}$ & $\tilde{\Pi}_{s}$   & $\tilde{\Pi}_{ss}$\\
  \hline
1&   $3520$ & $5.0{\times}10^{-3}$ & $9$ & $ 0 $                     & $1.0$ & $-5.0{\times}10^{-1}$ & $ 3.3 $  & $1.2{\times}10^{-2}$ & $ 3.3 $ & $2.5{\times}10^{-2}$\\
2&  $1710$ & $2{\times}10^{-2}$ & $0.8$ & $ 0 $                   & $1.0$ & $-7{\times}10^{-1}$ & $ 1.8 $ & $0$ & $0.47$  & $-3.9{\times}10^{-2}$\\
  \hline
3&  $390 $ & $8.5{\times}10^{-2}$ & $8.3$ & $3.4{\times}10^{-3}$ & $1.0$ & $ 7{\times}10^{-3}$ & $ 21 $ & $0$  & $2.1$  & $-0.83$\\
4& $125 $ & $0.5$ & $2.8$ & $2.8{\times}10^{-3}$ & $1.0$ & $ 9.7{\times}10^{-2}$ & $ 26 $ & $2.1$  & $ 9.0 $  & $2.3$ \\
  \hline \hline\\
\end{tabular}
}
\end{table*}

The multimode nonlinear BO model for liquid water consists of four primary oscillator modes representing (1) intramolecular OH stretching (``stretching''), (2) intramolecular HOH bending (``bending''), (3) HB--intermolecular librational (``librational''), and (4) HB--intermolecular translational (``translational'') motions.\cite{IT16JCP,TT23JCP1} Note that the OH stretching modes consist of symmetry and anti-symmetry modes, but here we treat the stretching modes as a single mode because we construct the model based on the broadened linear and nonlinear spectra as benchmarks, as in previous studies. Although it is possible to model the two modes separately by MD simulation with the machine learning approach,\cite{UT20JCTC} solving such a model is computationally demanding, so here we treat them as one mode;
thus, they are described by dimensionless vibrational coordinates $\vec{q}=(\dots,q_{s},\dots)$ with $s=1,\dots,4$ indexing the four vibrational modes. The total Hamiltonian is then expressed as
\begin{align}
{H}&=\sum_{s}\left( {H}^{(s)}_{\mathrm{S}}+{H}^{(s)}_{\mathrm{B}}+{H}^{(s)}_{\mathrm{I}} \right) \nonumber \\ &+ \frac{1}{2} \sum_{s>s'} \left( g_{s^2s'}q_{s}^2q_{s'} + g_{s{s'}^2}q_{s}q_{s'}^2 \right),
  \label{eq:h_total}
\end{align}
where
\begin{align}
{H}^{(s)}_{\mathrm{S}}=\frac{{p}_s^{2}}{2m_s}+\frac{1}{2!}m_{s}\omega_{s}^2q_{s}^2 + \frac{1}{3!}g_{s^3}q_{s}^3
  \label{eq:h_system}
\end{align}
is the Hamiltonian for the $s$th mode with mass $m_s$, coordinate ${q_s}$, momentum ${p_s}$, fundamental frequency $\omega_{s}$, and anharmonicity of potential $g_{s^3}$, and $g_{s^2s'}$ and $g_{s{s'}^2}$ are the anharmonic coupling between modes $s$ and $ s'$,
\begin{align}
  {H}^{(s)}_{\mathrm{B}}=\sum _{j_s}\left(\frac{{p}_{j_s}^{2}}{2m_{j_s}}+\frac{m_{j_s}\omega _{j_s}^{2}{x}_{j_s}^{2}}{2} \right)+ \sum _{j_s}\left(\frac{\alpha _{j}^{2}V_s^{2}({q_s})}{2m_{j_s}\omega _{j_s}^{2}}\right)
  \label{eq:h_bath}
\end{align}
is the bath Hamiltonian for the $s$th mode with the momentum, coordinate, mass, and frequency of the $j_s$th bath oscillator given by ${p}_{j_s}$, ${x}_{j_{s}}$, $m_{j_{s}}$ and
$\omega _{{j_s}}$, respectively, and
\begin{align}
  {H}^{(s)}_{\mathrm{I}}&=- V_{s}({q_s})\sum _{j_s}\alpha _{j_s}{x}_{j_s}
  \label{eq:h_int}
\end{align}
is the SB interaction, which consists of the linear--linear (LL)
and square--linear (SL) SB interactions, and
$V_{s}({q_s})\equiv V^{(s)}_{\mathrm{LL}}{q_s}+V^{(s)}_{\mathrm{SL}}{q_s}^{2}/2$, with coupling strengths of $V^{(s)}_{\mathrm{LL}}$, $V^{(s)}_{\mathrm{SL}}$,
and $\alpha _{j_s}$.\cite{OT97PRE,TS20JPSJ,KT04JCP,IT06JCP,ST11JPCA,IIT15JCP}
The last term of the bath Hamiltonian is a counter-term that maintains the translational symmetry of the system.\cite{T06JPSJ}
While the LL interaction contributes mainly to energy relaxation, the SL system--bath interaction leads to vibrational dephasing in the slow modulation case due to the frequency fluctuation of system vibrations.\cite{IT06JCP,ST11JPCA,IIT15JCP}
The sum of the bath coordinates, ${X}_s\equiv \sum _{j_{s}}\alpha _{{j_s}}{x}_{{j_s}}$, acts as a collective coordinate that modulates mode $s$.\cite{TK89JPSJ1,T06JPSJ,T20JCP}
We introduce the spectral distribution function (SDF)
$J_s(\omega )\equiv \sum _{j_s}\alpha _{j_s}^{2}\hbar \delta (\omega -\omega _{j_s})/2m_{j_s}\omega _{j_s}$ to characterize the environmental fluctuation.
We assume that $J_s(\omega )$ has an Ohmic form with a Lorentzian cutoff:
\begin{align}
  J_s(\omega )=\frac{\hbar m_s \zeta_s }{2\pi }\frac{\omega \gamma_s ^{2}}{\gamma_s^{2}+\omega ^{2}}.
\end{align}
Here, $\zeta_s$ is the system--bath coupling strength, and $\gamma_s$ represents the width of the spectral distribution for mode $s$ that relates to the vibrational dephasing time, defined as $\tau_s=1/\gamma_s$.

\begin{table}[!tb]
\caption{\label{tab:FitAll2}Parameter values of multimode LL+SL BO model for anharmonic mode--mode coupling and optical properties among (1) stretching, (2) bending, (3) librational, and (4) translational modes on the basis of Refs. \onlinecite{IT16JCP,TT23JCP1}. The normalized parameters are defined as $\tilde{g}_{s^2s'}\equiv (\omega_0^3/\omega_s^2 \omega_s') g_{s^2s'}$, $\tilde{g}_{s{s'}^2}\equiv (\omega_0^3/\omega_s \omega_{s'}^2) g_{s{s'}^2}$,
$\tilde{\mu}_{ss'} \equiv (\omega_0^2/\omega_s \omega_s') \mu_{ss'}$, and $\tilde{\Pi}_{ss'} \equiv  (\omega_0^2/\omega_s \omega_s') \Pi_{ss'}$.
}
\begin{tabular}{cccccccccccc}
  \hline \hline
  $\mathrm{s-s'}$ & $\tilde{g}_{s^2s'}$ & $\tilde{g}_{s{s'}^2}$ & $\tilde{\mu}_{ss'}$ & $\tilde{\Pi}_{ss'}$ \\
  \hline
  $\mathrm{1-2}$  & $0$ & $0.2$ & $2.0 \times 10^{-3}$ & $2.6 \times 10^{-3}$ \\
  $\mathrm{1-3}$  & $-3.9 \times 10^{-2}$ & $-3.9 \times 10^{-2}$ & $0.13$ & $0.19$ \\
  $\mathrm{1-4}$  & $-7.5 \times 10^{-2}$ & $-7.5 \times 10^{-2}$ & $0.43$ & $0.46$ \\
  $\mathrm{2-3}$  & $-1.5 \times 10^{-2}$ & $-1.5 \times 10^{-2}$ & $7.0$ & $4.0$ \\
  $\mathrm{2-4}$  & $-2.0 \times 10^{-2}$ & $-2.0 \times 10^{-2}$ & $3.1\times 10^{-2}$ & $3.1 \times 10^{-2}$ \\
  $\mathrm{3-4}$  & $0.23$ & $0.23$ & $7.8 \times 10^{-2}$ & $0.16$ \\
  \hline \hline\\
\end{tabular}
\end{table}

The dipole operator and polarizability for the $s$th mode are defined as
\begin{equation}
  \hat{\mu}_s =  \mu_s {q}_s + \sum_{s'} \mu_{ss'} {q}_s {q}_{s'}
\label{eq:mu}
\end{equation}
and 
\begin{equation}
  \hat{\Pi}_s =  \Pi_s {q}_s + \sum_{s'} \Pi_{ss'} {q}_s {q}_{s'},
\label{Pi}
\end{equation}
where $\mu_s$ and $\mu_{ss'}$ are the linear and nonlinear elements of the dipole moment, and $\Pi_s$ and $\Pi_{ss'}$ are those of the polarizability, respectively. The vibrational modes interact through the mechanical anharmonic coupling (MAHC) described by ${g}_{s^2 s'}$ and ${g}_{ss'^2}$ and the electric anharmonic coupling (EAHC) described by ${\mu}_{ss'}$ and ${\Pi}_{ss'}$. Although the effects of MAHC and EAHC in 2D spectra have been discussed analytically,\cite{OT03JPC,Hamm2DTHzFeynmann,OT97JCP1,OT97JCP2,OT97JCP2,OT97CPL2,OJT01CP} no quantum theory exists for complex systems involving complex system--bath interactions, such as those in this study.

In the present case, the difference between the fundamental frequencies of the intermolecular and intramolecular modes is very large, and we can excite each mode separately by choosing the light sources: a high-frequency intramolecular mode is solely excited by an IR pulse, whereas a low-frequency intermolecular mode is solely excited by a Raman or THz pulse. Note that, although the contribution from EAHC arises from nonlinear dipole elements or nonlinear polarizability elements are not important for a spectroscopic approach based on the third-order response functions for $s=$ 1 and 2 and they can be ignored,\cite{OT03JPC,Hamm2DTHzFeynmann} we used the same parameter values presented in Ref. \onlinecite{TT23JCP1} as given in Tables \ref{tab:FitAll1} and \ref{tab:FitAll2} because the computational cost in our approach is the same.
Besides these, we used the same parameter values presented in Ref. \onlinecite{TT23JCP1}. The element $\mu_{23}$ is important even in 1D IR because this is the cause of the combination band at the frequency $\omega=2100$ cm$^{-1}$.

We chose a set of parameters for the model based on the MD results of 2D spectra and experimentally obtained 1D spectra, taking advantage of the fact that the 2D profiles differ significantly for different dynamic processes.\cite{IT16JCP,TT23JCP1}

\section{Calculating 2D correlation spectra of third-order IR and fifth-order IR--Raman measurements}\label{sub:HEOM2D}
\begin{figure*}[htbp]
  \centering
  \includegraphics[keepaspectratio, scale=0.09]{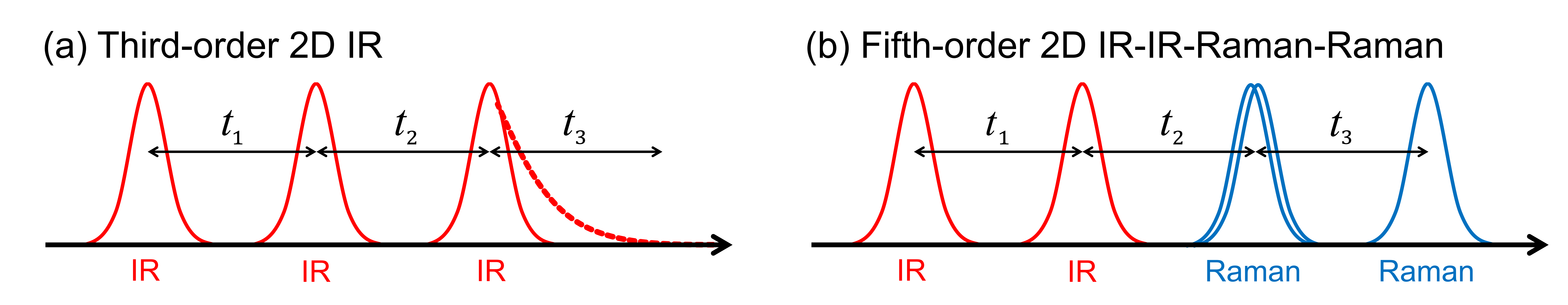}
  \caption{Pulse sequence of third-order 2D IR and fifth-order 2D IR--IR--Raman--Raman spectroscopies.}
  \label{fgr:pulse}
\end{figure*}
We consider 2D IR and IR--Raman spectroscopies whose pulse sequences are depicted in Fig. \ref{fgr:pulse}. These spectra can be calculated from the three-body nonlinear response function defined as\cite{T06JPSJ,T20JCP}
\begin{equation}
R(t_3,t_2,t_1) = \qty(\frac{i}{\hbar })^{3}\mathrm{tr}\qty{\hat{A}\mathcal{G}(t_{3})\hat{B}^{\times}\mathcal{G}(t_{2})\hat{C}^{\times }\mathcal{G}(t_{1})\hat{D}^{\times }\hat{\rho }^{\mathrm{eq}}},
  \label{eq:RFTIRI}
\end{equation}
where the operators $\hat{A}$, $\hat{B}$, $\hat{C}$, and $\hat{D}$ are either the dipole moment $\hat \mu$ or the polarizability $\hat \Pi$. Here, we have employed the hyperoperator $^{\times }$ defined as $\hat{A}^{\times }\hat{{\hat \rho}}\equiv [\hat{{A}},{\hat \rho}]$, $\mathcal{G}(t)$ is the Green's function of the total Hamiltonian without a laser interaction, and $\hat{\rho }_{\mathrm{eq}}$ is the equilibrium state. Using this expression, the third-order IR signal is calculated from $\hat A$$\sim$$\hat D=\hat{\mu}$
and the fifth-order IR--Raman signal is calculated from $\hat{A}=
\hat{B}=\hat{\Pi}$ and $\hat{C}=\hat{D}=\hat{\mu}$, respectively.\cite{TS20JPSJ,KT04JCP,IT07JPCA} Note that, in our model calculation, the role of IR and Raman excitation for the intramolecular modes and that of THz and Raman excitation for the intermolecular modes are similar because the nonlinear elements of each mode, such as $\mu_{ss}$ and $\Pi_{ss}$, do not play a major role in determining the profiles of the 2D spectrum, which is defined as Eq. \eqref{eq:RFTIRI}. This is because nonlinear spectroscopy based on a three-body response function such as $R_{\mathrm{IIR}}^{(3)}(t_{2},t_{1})=-\langle [[\hat{\bm{\Pi }}(t_{2}+t_{1}),\hat{\bm{\mu }}(t_{1})],\hat{\bm{\mu }}(0)]\rangle /\hbar ^{2}$
is described as an odd number of optical operators, whereas that based on a four-body response function such as $R_{\mathrm{IIRR}}^{(5)}(t_{3},t_{2},t_{1})=-i\langle [[\hat{\bm{\Pi}}( t_{3}+t_{2}+t_{1}), [[\hat{\bm{\Pi }}(t_{2}+t_{1}),\hat{\bm{\mu }}(t_{1})],\hat{\bm{\mu }}(0)]]\rangle /\hbar ^{3}$ is described as an even number of optical operators.  When the anharmonicity of the mode described by $g_{s^3} q_s^3$ is weak, the thermal statistical average included in the calculation of the response function is performed as a Gaussian integral of $q_s$. Therefore, $R_{\mathrm{IIR}}^{(3)}(t_{2},t_{1})$ must be of even order function of $q_s$ so that it does not disappear in the Gaussian integral, and a contribution of ${\Pi}_{ss}$ or ${\mu}_{ss}$ is required. In contrast, $R_{\mathrm{IIRR}}^{(5)}(t_{3},t_{2},t_{1})$ is fourth order in $q_s$ even without ${\Pi}_{ss}$ or ${\Pi}_{ss}$, and, if they contribute, it should be of the second order in ${\Pi}_{ss}$ and/or ${\mu}_{ss}$, such as ${\Pi}_{ss}{\mu}_{ss}$.\cite{TM93JCP}
This indicates that the MAHC contribution, which is often referred as the non-Condon effect,\cite{Skinner_nonCondon2005} in $R_{\mathrm{IIR}}^{(3)}(t_{2},t_{1})$ and $R_{\mathrm{IIRR}}^{(5)}(t_{3},t_{2},t_{1})$ measurements is minor. Thus, although it would be difficult to conduct an actual measurement, the profile of a seventh-order 2D Raman spectrum (i.e., $\hat A$$\sim$$\hat D=\hat{\Pi}$)\cite{TS20JPSJ,KT04JCP,PBSLTM94JPC} should be similar to that of third-order 2D IR. 

The calculated signals are plotted as the 2D correlation spectrum that is obtained by adding the two terms corresponding to the rephasing and nonrephasing parts of $R(t_3,t_2,t_1)$ in equal weights.
A commonly used definition of a 2D correlation spectrum is expressed as\cite{2DCrrJonas2001,2DCrrGe2002,2DCrrTokmakoff2003} 
\begin{align}
&I_{\rm{C}}({\omega _3},{t_2},{\omega _1}) \nonumber \\ 
&= {\rm Im} \left\{ \int_0^\infty  \dd {t_1} \int_0^\infty  \dd {t_3}
 e^{ - i{\omega _1} {t_1} }  e^{ + i{\omega _3}{t_3}} 
 R_{\rm{R}} ({t_3},{t_2},{t_1})\right\}  \nonumber \\ 
&  + {\rm Im} \left\{ \int_0^\infty  \dd{t_1}  \int_0^\infty  \dd{t_3} {e^{i{\omega _1}{t_1}}}{e^{ + i{\omega _3}{t_3}}} R_{{\rm{NR}}}({t_3},{t_2},{t_1}) \right\}, \nonumber \\
  \label{2DCorr}
\end{align}
where $R_{{\rm{R}}}$ and $R_{{\rm{NR}}}$ are the rephasing and nonrephasing responses that are experimentally detected separately using the phase-matching conditions. Theoretically, although we can separate $R_{{\rm{R}}}$ and $R_{{\rm{NR}}}$ by choosing the specific Liouville paths in the energy state model, the process is not easy when we calculate the response functions in the coordinate space. Therefore, here we eliminate the undesired rephasing contribution by utilizing the Fourier transformation of $t_2$ for $I_{\rm{C}}({\omega_3},{t_2},{\omega _1})$ with $R_{\rm{R}} ({t_3},{t_2},{t_1})=
 R_{{\rm{NR}}}({t_3},{t_2},{t_1})=R(t_3,t_2,t_1)$ to remove the oscillating contribution in period $t_2$ period with frequency 
 $2\omega_s$, where $\omega_s$ is the frequency of the targeting mode $s$.\cite{HT08JCP,YagasakiSaitoJCP20082DIR} 
Note that we can also separate the plain response function in terms of different phase-matching conditions using a phase-matching transformation,\cite{KT01CPL} but the Fourier based method is simpler when only a 2D correlation spectrum is calculated.

After removing the dispersive contribution of the response function, we obtain the pure absorptive line shape. We observe a distorted line shape, called the phase-twisted line, which exhibits the non-Markovian and nonperturbative dynamics of the environment.
\cite{TI09ACR,Cho2009,Hamm2011ConceptsAM,2DCrrJonas2001,2DCrrGe2002,2DCrrTokmakoff2003} To investigate not only anharmonic mode--mode coupling but also vibrational dephasing and thermal relaxation effects at finite temperature, we must employ nonperturbative and non-Markovian open quantum dynamics theory, which enables us to treat the quantum entanglement between the system and bath.\cite{T20JCP} 
Therefore, we employ the discretized hierarchical equations of motion in mixed Liouville--Wigner space (DHEOM--MLWS): Lagrange--Hermite mesh discretization is employed in the Liouville space of the intramolecular modes, and Lagrange--Hermite mesh discretization and Hermite discretization are employed in the Wigner space of the intermolecular modes.\cite{TT23JCP1} 

\section{RESULTS AND DISCUSSION}
\label{result}
Although the DHEOM--MLWS for the water BO model was created to treat the above four modes simultaneously, the computational time required for 2D vibrational spectra is about 1000 times longer than that for linear spectra, even with a fixed $t_2$. Therefore, here we calculate 2D spectra for mode--mode coupling, which can be investigated for two modes, as in combinations of stretching--bending (1--2) and bending--librational (2--3) for 2D correlation IR spectra, and stretching--librational (1--3) and stretching--translational (1--4) for 2D correlation IR--Raman spectra. To calculate the spectrum, we integrate the DHEOM--MWLS using the same numerical setup as in Ref. \onlinecite{TT23JCP1} with the parameter set presented in Tables \ref{tab:FitAll1} and \ref{tab:FitAll2}. Thus, the calculated 1D IR and 1D Raman spectra (not shown) are identical to those presented in Ref. \onlinecite{TT23JCP1}.

Note that the dipole moment and polarizability have similar functional forms in this model calculation. Therefore, if the same modes are excited, 2D Raman--Raman--IR--IR or 2D IR--Raman--IR--Raman exhibit a profile that is similar to the present 2D IR--IR--Raman--Raman results. 

\subsection{2D correlation IR spectra}
\subsubsection{Stretching--bending modes}
\begin{figure*}[!t]
  \centering
  \includegraphics[scale=0.45]{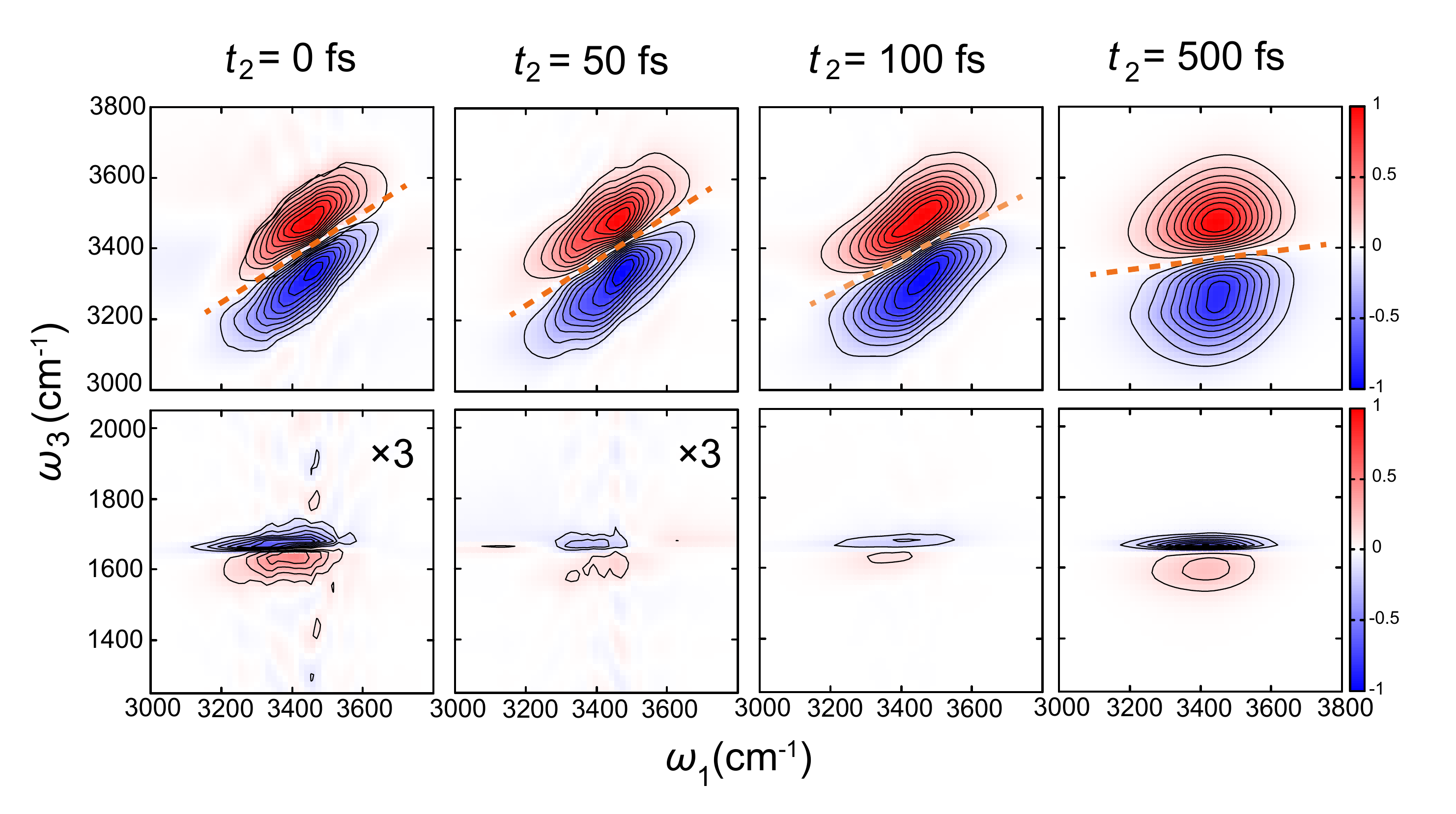}
  \caption{Third-order 2D correlation IR spectra for stretching motion (upper panel) and stretching $\rightarrow$ bending motion (lower panel) for different $t_2$. All spectral intensities are normalized with respect to the absolute value of the maximum peak intensity of each diagonal peak. The direction of the nodal lines (orange dashed lines) in the upper panel represents the extent of correlation between the vibrational coherences of the $t_1$ and $t_3$ periods. For clarity, the data of off-diagonal peaks are multiplied by 3 in the cases of $t_2=0$ fs and $50$ fs.}
  \label{fgr:2DIRA}
\end{figure*}
Figure \ref{fgr:2DIRA} illustrates 2D correlation spectra calculated for the stretching--bending (1--2) modes for which experimental results are available.\cite{Tokmakoff2016H2O,VothTokmakoff_St-BendJCP2017,Tokmakoff2022} 
Note that previous theoretical investigations for water 2D spectroscopy have been limited to stretching modes only for quantum calculations based on classical MD\cite{SkinnerStochs2003,Mukamel2009,JansenSkiner2010} and individual stretching and bending modes for classical MD.\cite{ImotXanteasSaitoJCP2013H2O,ChoH2OMD2014} The present investigations provide the first quantum simulation of 2D spectra for intramolecular--intramolecular and intramolecular--intermolecular mode--mode couplings.
To conduct the simulations, we set $\hat{\bm{\mu}}=\hat{{\mu}}_1+\hat{{\mu}}_2$.
Each peak near $(\omega_1,\omega_3)=(3400 \rm {cm}^{-1}, 3400 cm^{-1})$ in the upper panel represents the stretching mode; the red positive peaks and the blue negative peaks correspond to the transitions between the $|0\rangle_s$$\rightarrow_s$$|1\rangle_s$$\rightarrow$$|0\rangle_s$ and $|0\rangle_s$$\rightarrow$$|1\rangle_s$$\rightarrow$$|2\rangle_s$ states for $s=1$, respectively, where $|n\rangle_s$ is the $n$th vibrational energy state of the $s$th mode.

In the lower panel of each figure in Fig. \ref{fgr:2DIRA} present the cross peaks representing the stretching$\rightarrow$bending transition (e.g., $|0\rangle_1 |0\rangle_2$$\rightarrow$$|1\rangle_1 |0\rangle_2$$\sim$$|0\rangle_1 |1\rangle_2$$\rightarrow$$|0\rangle_1 |0\rangle_2$).\cite{TI09ACR} 
The coupling peaks that appear at $t_2=0$ are due to coherent transfer between the two modes and quickly disappear at $t_2=50$fs, when coherence is lost. The coupling peaks that appear after $t_2\ge$100 fs are due to population transfer, and their intensity increases with $t_2$. 

The direction of the orange nodal lines represents the extent of noise correlation (non-Markovian effects) between the vibrational coherences of periods $t_1$ and $t_3$. The direction parallel to the $\omega_1$ axis is the uncorrelated case, whereas the direction parallel to the $\omega_1=\omega_3$ line is the fully correlated case.\cite{2DCrrJonas2001,2DCrrGe2002,2DCrrTokmakoff2003} The width of the peak parallel to the $\omega_1=\omega_3$ direction corresponds to inhomogeneous broadening, whereas that perpendicular to $\omega_1=\omega_3$ is homogeneous broadening.\cite{Wiersma2006,TI09ACR} 

The spectral width of the stretching peaks in the upper panel of Fig. \ref{fgr:2DIRA} is broadened due to the dephasing (spectral diffusion regime) that arises from the non-perturbative and non-Markovian effects of the SL interaction.\cite{IT06JCP} When $t_2$ becomes larger than the correlation time of the noise $\tau_1=1/\gamma_1\approx 160$ fs, the correlation begins to weaken, and at $t_2=500$ fs, the nodal line becomes horizontal. The SL interaction used in this model causes the same effect as in stochastic theory, where $\tau_1$ determines the dephasing time.\cite{T06JPSJ} We found that our result is consistent with those of previous experiments \cite{Bakker1999,ElsaesserDwaynePNAS2008,Tokmakoff2016H2O,VothTokmakoff_St-BendJCP2017,Tokmakoff2022} and theoretical calculations.\cite{SkinnerStochs2003,Mukamel2009,JansenSkiner2010,ImotXanteasSaitoJCP2013H2O,ChoH2OMD2014}

Upon increasing $t_2$, it was observed experimentally that the blue 0--1--2 peak extended in the small $\omega_3$ direction for $t_2 \le 250$fs, and a peak, referred to as the hot ground state, appeared above the red 0--1--0 peak for $t_2\ge 250$fs. \cite{ElsaesserDwaynePNAS2008,Tokmakoff2016H2O,VothTokmakoff_St-BendJCP2017,Tokmakoff2022} In contrast, our results show that the blue peak extends only slightly in the negative $\omega_3$ for $t_2 \ge 500$fs direction due to the population relaxation arising from the SL interaction,\cite{IT06JCP} while no hot ground state peak is observed. Note that the inclusion of the LL interactions in the stretching\cite{IT06JCP,ST11JPCA} and/or the presence of non-radiative decay states\cite{UT20JCTC} would enhance the elongation. To predict such phenomena with a current handy model calculation, one must include various hypothetical factors that are considered important in determining the vibrational process of water, with reference to the experimentally obtained 2D spectral profile. We have computed 2D spectra for the case of including quartic Darling--Dennison (DD) coupling described as $\tilde{g}_{1^22^2}q_1^2 g_2^2$.\cite{OJT01CP,TI09ACR} and for the case of stronger cubic (Fermi) coupling  ($\tilde{g}_{s{s'}^2}=0.5$). While inclusion of the DD coupling does not change the 2D profile much (not shown), we found that, when the Fermi coupling is strong, the 2D profile is relatively close to the experimental results (see Appendix \ref{sec:2DIRA}). However, some other information, such as quantum MD results, is needed to explore such a possibility in greater detail.

The cross peaks representing the stretching$\rightarrow$bending transition in the lower panel of Fig. \ref{fgr:2DIRA} 
are uncorrelated due to the process involved in the bending mode having a very short noise correlation time, $1/\gamma_2\approx 40$fs, and the positive and negative peaks are elongated in the $\omega_1$ direction. \cite{Tokmakoff2016H2O,VothTokmakoff_St-BendJCP2017,Tokmakoff2022} In Appendix \ref{sec:2DIRB}, we depict the 2D spectrum of the bending mode. It can be seen that the time scale of the spectral diffusion of the cross-peak is determined by the mode in which it is faster.

\subsubsection{Bending--librational modes}
\begin{figure}[htbp]
  \centering
  \includegraphics[keepaspectratio, scale=0.4]{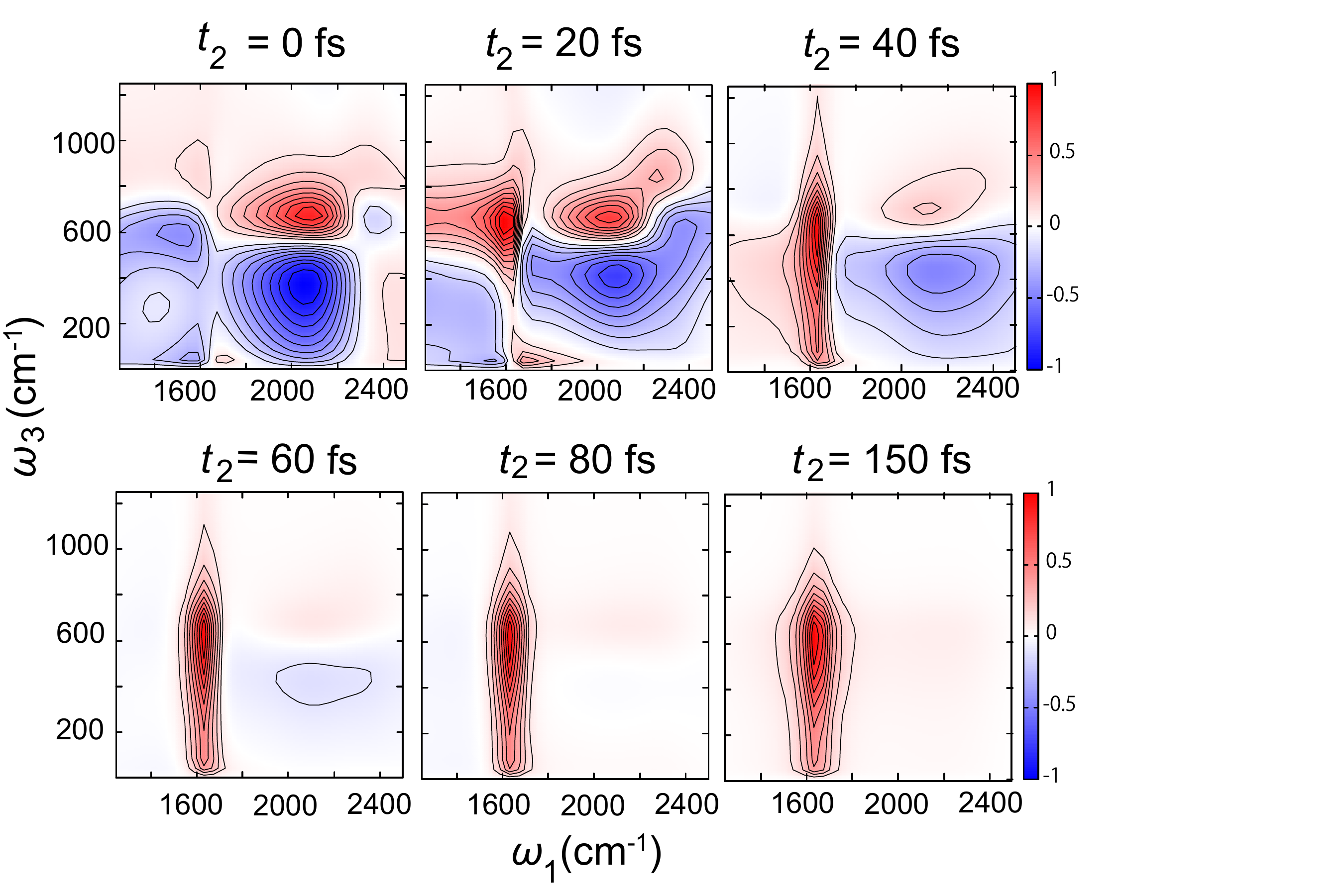}
  \caption{Third-order 2D correlation IR spectra for the bending$\rightarrow$librational (2$\rightarrow$3) transition for different $t_2$. All spectral intensities are normalized with respect to the absolute value of the maximum peak intensity in the $t_2=0$ case.}
  \label{fgr:bendlib}
\end{figure}
Next, we present the calculated result for the bending$\rightarrow$librational (2$\rightarrow$3) transition peak in Fig. \ref{fgr:bendlib}. We chose the first two IR pulses for the excitation of the bending mode and last two IR (or THz) pulses for the detection of the librational modes. In the 2D correlation spectrum based on the third-order (four body) response function, we can interpret the 2D spectral profile as investigating the excitation of each vibrational mode in the $\omega_1$ direction in the IR region and then detecting the dynamics of that excited motion after $t_2$ as the absorption spectrum in the $\omega_3$ direction in the THz region. In Appendix \ref{sec:2DIRB}, we depict the 2D spectrum of the bending mode for the same conditions as in Fig. \ref{fgr:bendlib}.

In the present case, the first two IR pulses excite the bending mode ($\omega_1=1600$ cm$^{-1}$) and the combination band (around $\omega_1=2100$ cm$^{-1}$) through $\mu_2$ and $\mu_{23}$, respectively. Because no population transfer for 2$\rightarrow$3 occurs at time $t_2=0$, we cannot observe a prominent peak in the $\omega_3$ direction along the line $\omega_1=1600$ cm$^{-1}$. The positive and negative peaks of the combination band peak around $\omega_1=2100$ cm$^{-1}$ appear because the IR excitations described by $\mu_{23}$ cause transitions of the librational vibrational states, $|0\rangle_2 |0\rangle_3$$\rightarrow$$|1\rangle_2|1\rangle_3$$\rightarrow$$|1\rangle_2 |0\rangle_3$ and $|0\rangle_2 |0\rangle_3$$\rightarrow$$|1\rangle_2|1\rangle_3$$\rightarrow$$|1\rangle_2 |2\rangle_3$,
which appear as the positive upper and negative lower peaks, respectively.

At $t_2=20$ fs, the elongated peak along line $\omega_1=1600$ cm$^{-1}$ appears as the result of the population transfer from 2$\rightarrow$3. The elongation of the peak along this line reflects the fact that the librational mode is spread out. Because of this, the positive and negative combination band peaks are also elongated in the $\omega_3$ direction. The nodal line of the combination band peaks in the high-frequency region is twisted, indicating that the contribution from the librational mode, which seems to dominate the high-frequency part of the combination band, can be correlated with the IR excitation of the librational mode. Because these combination band peaks appear due to the coherent process $|0\rangle_2$$\rightarrow$$|1\rangle_2$, they decay quickly and almost disappear at $t_2=80$ fs, whereas the 2$\rightarrow$3 transition peak along line $\omega_1=1600$cm$^{-1}$ grows because this process arises from the relaxation. 
The fast relaxation found between these two modes is also consistent with other classical MD analyses.\cite{YagasakiSaitoJCP2011Relax} 

\subsection{Fifth-order 2D correlation IR--IR--Raman--Raman spectra}
Next, we display our calculated results for the mode--mode coupling of the high-frequency stretching mode to the low-frequency intramolecular modes. Here, we chose IR pulses for the excitation of the stretching mode and Raman pulses for the detection of the intermolecular modes because we believe that such an experimental setup is feasible with current technology. However, because the difference in the mathematical treatment of Raman and IR (or THz) excitation is minor in our model calculation, the spectral profile would be similar regardless of the laser configuration, for example, for a configuration where Raman pulses excite the stretching mode and THz pulses detect the intermolecular modes.

\subsubsection{Stretching--translational modes}
\begin{figure}[htbp]
  \centering
  \includegraphics[keepaspectratio, scale=0.43]{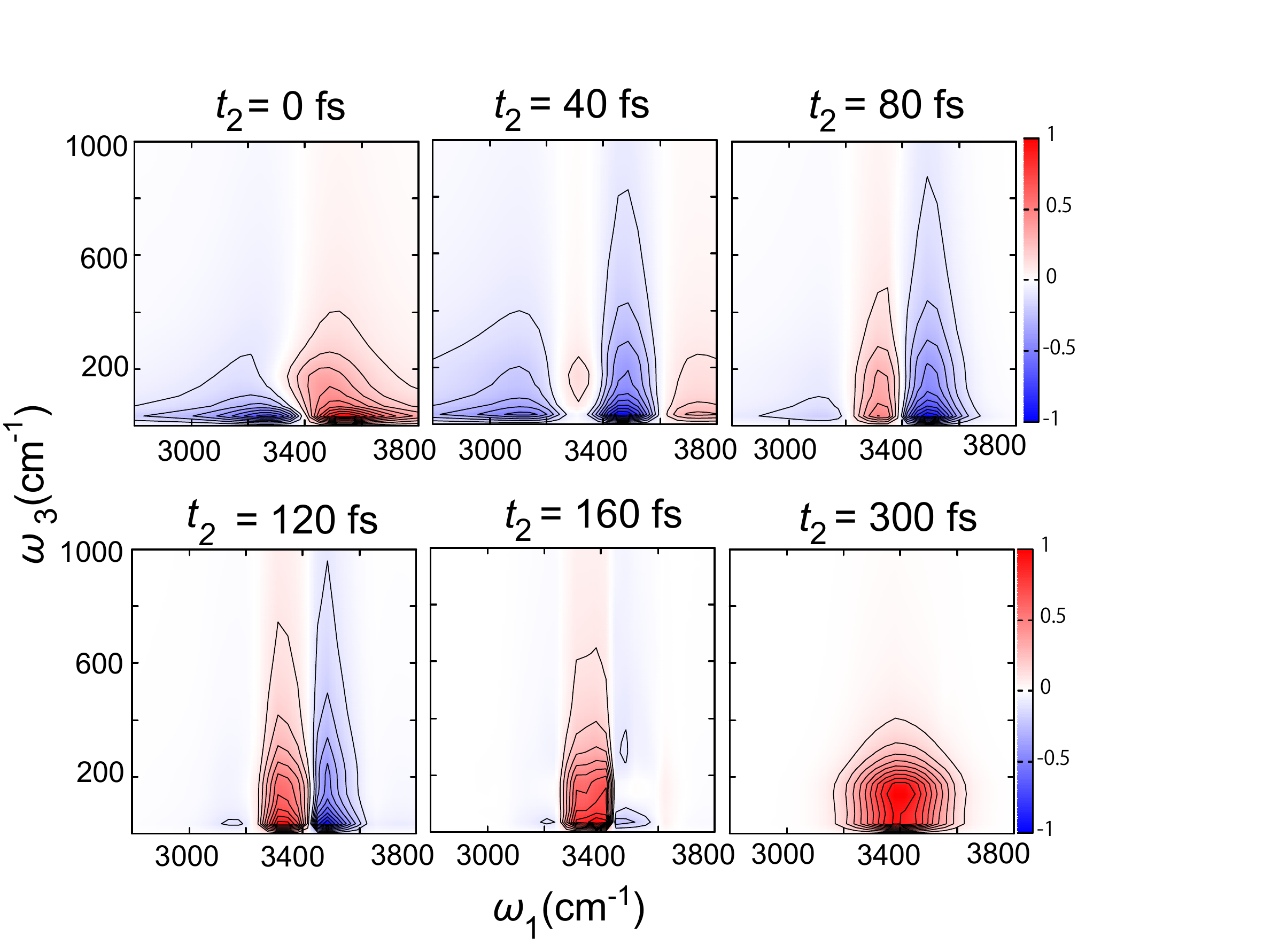}
  \caption{Fifth-order 2D correlation IR--IR--Raman--Raman spectra for the stretching$\rightarrow$translational (1$\rightarrow$4) coupling for different $t_2$. All spectral intensities are normalized with respect to the absolute value of the maximum peak intensity in the $t_2=0$ case.}
  \label{fgr:strtran}
\end{figure}
We now present the simulation result of fifth-order 2D IR--IR--Raman--Raman spectra for the stretching--translational coupling case. Note that this process was investigated theoretically and experimentally through the use of 2D THz--IR--visible (or 2D IR--IR--Raman) measurement.\cite{IT16JCP,grechko2018,Bonn2DTZIFvis2021,TT23JCP1} 
The cross peaks predicted from the 2D THz--IR--visible spectra arise from both MAHC and EAHC contributions of the stretching--translational modes, but, in the 2D IR--IR--Raman--Raman spectra, the cross peaks always arise from MAHC, whereas those from EHAC appear at different positions or do not appear, as can be seen from the analytical expressions of the response functions.\cite{OT97JCP1,OT97JCP2,OT97CPL2,OT03JPC,Hamm2DTHzFeynmann}  

Moreover, although 2D THz--IR--visible measurement is designed to detect the coherence between the two different modes through the mode--mode interactions, the present 2D IR--IR--Raman--Raman spectroscopy is designed to detect the population transfer among modes under dephasing and relaxation through the use of period $t_2$. Thus, the information obtained from the two approaches is different and complimentary.

The results of 2D IR--IR--Raman--Raman (or 2D correlation IR--IR--Raman--Raman) spectra for different $t_2$ values are presented in Fig. \ref{fgr:strtran}. As in the bending-librational case in Fig. \ref{fgr:bendlib}, the first two IR pulses applied at time duration $t_1$ excite not only the fast stretching mode but also the slow translational mode due to the presence of the large $\mu_{14}$ element. As a result, the phase of the stretching motion (with large $\omega_1$) is twisted by the slow translational motion.
Thus, the intensities of the stretching--translational peaks vary as a function of $t_2$. Because the vibrational motion of the translational mode decays quickly due to the large LL interaction, the phase twist disappears around $t_2=160$ fs, and a single positive peak is observed along line $\omega_1=3400$ cm$^{-1}$ at $t_2=300$ fs. This peak is elongated in the $\omega_3$ direction, reflecting the fact that the translational mode extends into the high-frequency region.
  
\subsubsection{Stretching--librational modes}
\begin{figure}[htbp]
  \centering
  \includegraphics[keepaspectratio, scale=0.46]{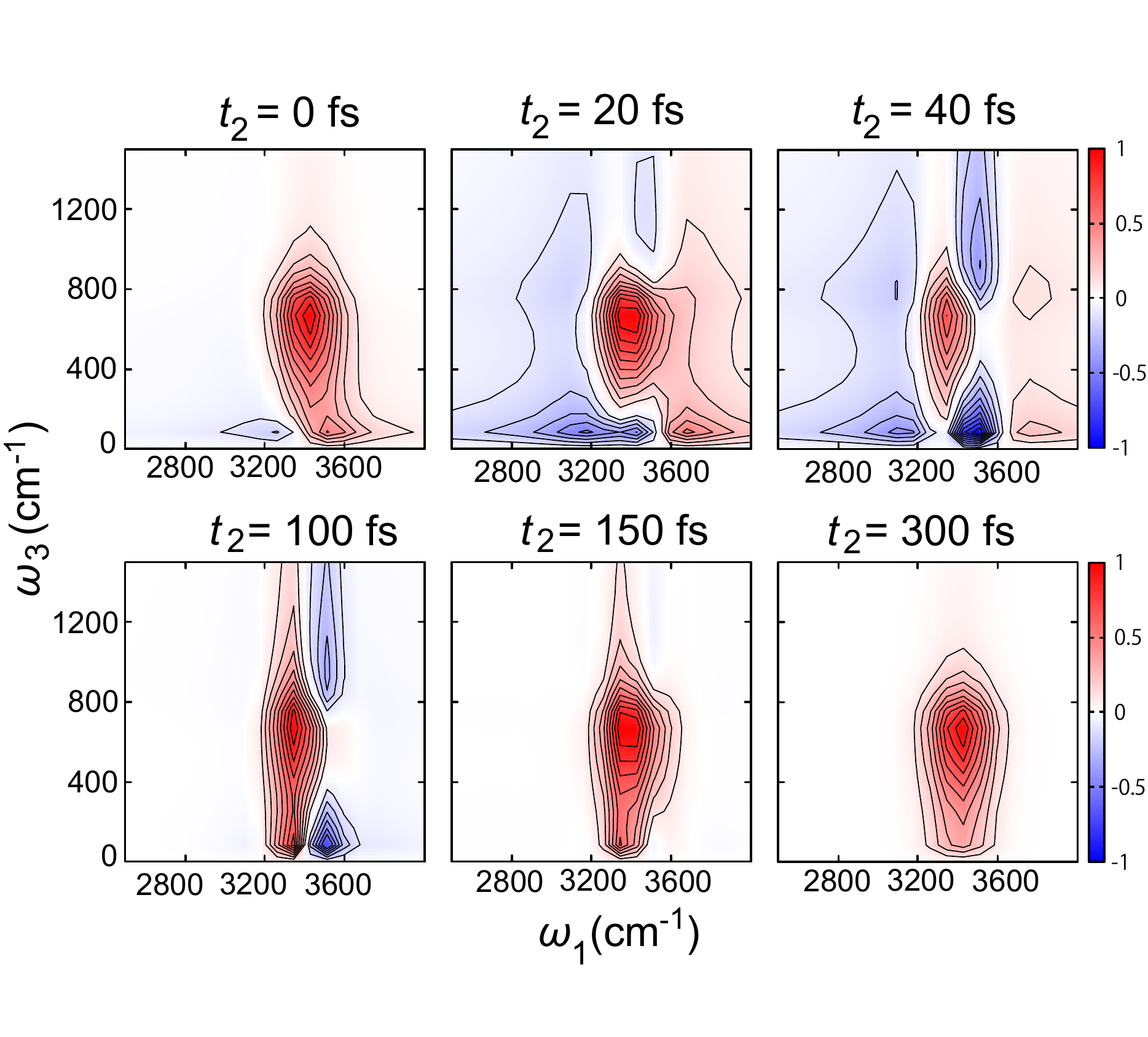}
  \caption{2D correlation IR--IR--Raman--Raman spectra for stretching$\rightarrow$librational ($1\rightarrow$3) coupling for different $t_2$ values. All spectral intensities are normalized with respect to the absolute value of the maximum peak intensity in the $t_2=0$ case.}
  \label{fgr:stlib}
\end{figure}
Finally, 2D correlation IR--IR--Raman--Raman spectra for the stretching--librational case are plotted in Fig. \ref{fgr:stlib}. As shown in Table \ref{tab:FitAll2}, the intensity of the $\mu_{13}$ element is comparable to $\mu_1$. Thus, the first two infrared pulses excite not only the stretching mode but also the librational mode, and, as in the stretching--translational case, the spectral profile changes in time rapidly for small $t_2$ values due to the phase twist. After librational oscillation decays around $t_2 =100$ fs, we observe a single positive peak at the stretching--librational coupling position, which extends in the $\omega_3$ direction due to the large broadening of the librational peak.

\section{CONCLUSION}\label{sec:conc}
We calculated third-order 2D correlation IR and fifth-order 2D correlation IR--IR--Raman--Raman spectra for the stretching--bending (1-2), bending--librational (2-3), stretching--librational (1-3), and stretching--translational (1-4) modes of liquid water using the DHEOM--MLWS approach based on the multimode LL BO model. Because the peaks that arise from EAHC play a minor role in four-body response function--based spectroscopies, we can easily discuss the role of mode--mode couplings that arise from MAHC processes. As the peaks arising from MAHC in this order of spectroscopy are associated with the population transfer process, they appear after the population relaxation occurs, whereas the peaks arising from EAHC are associated with the coherent transfer and appear at $t_2=0$ before decaying rapidly. 

To reproduce a 2D spectral profile, the model must capture the dynamical properties of the vibrational motions accurately. Thus, the scientific scenario underlying the model can be validated by calculating 2D spectra and comparing them to those obtained from experiments or accurate simulations. Taking advantage of the low computational cost and simplicity of the model, we can easily examine the mechanism to determine the 2D profile of spectrum, as briefly demonstrated in Appendix \ref{sec:2DIRA}. Although computationally expensive, within the framework of HEOM formalism, it is possible to describe the OH--stretching modes separately as symmetric and antisymmetric modes\cite{UT20JCTC} or to introduce another optically inactive mode that leads the excitation of the stretching mode to the vibrational ground state.\cite{IT18CP} However, the choice of models and model parameters should be justified by comparing results obtained by advanced experimental and simulation techniques.

With the limitation of CPU power, here we considered only the combination of the two-mode cases. To investigate a pathway of energy relaxation from the OH--Stretching mode to the low-frequency intramolecular modes, we should consider at least three modes, such as the stretching $\rightarrow$ bending $\rightarrow$ librational (1 $\rightarrow$ 2 $\rightarrow$ 3) modes or the stretching $\rightarrow$ bending $\rightarrow$ translational (1 $\rightarrow$ 2 $\rightarrow$ 4) modes including the anharmonicity of the three modes described as $\tilde{g}_{s{s'}{s''}}q_s q_{s'} q_{s''}$.  We leave such investigations to future studies. 

\begin{figure*}[!t]
  \centering
  \includegraphics[scale=0.45]{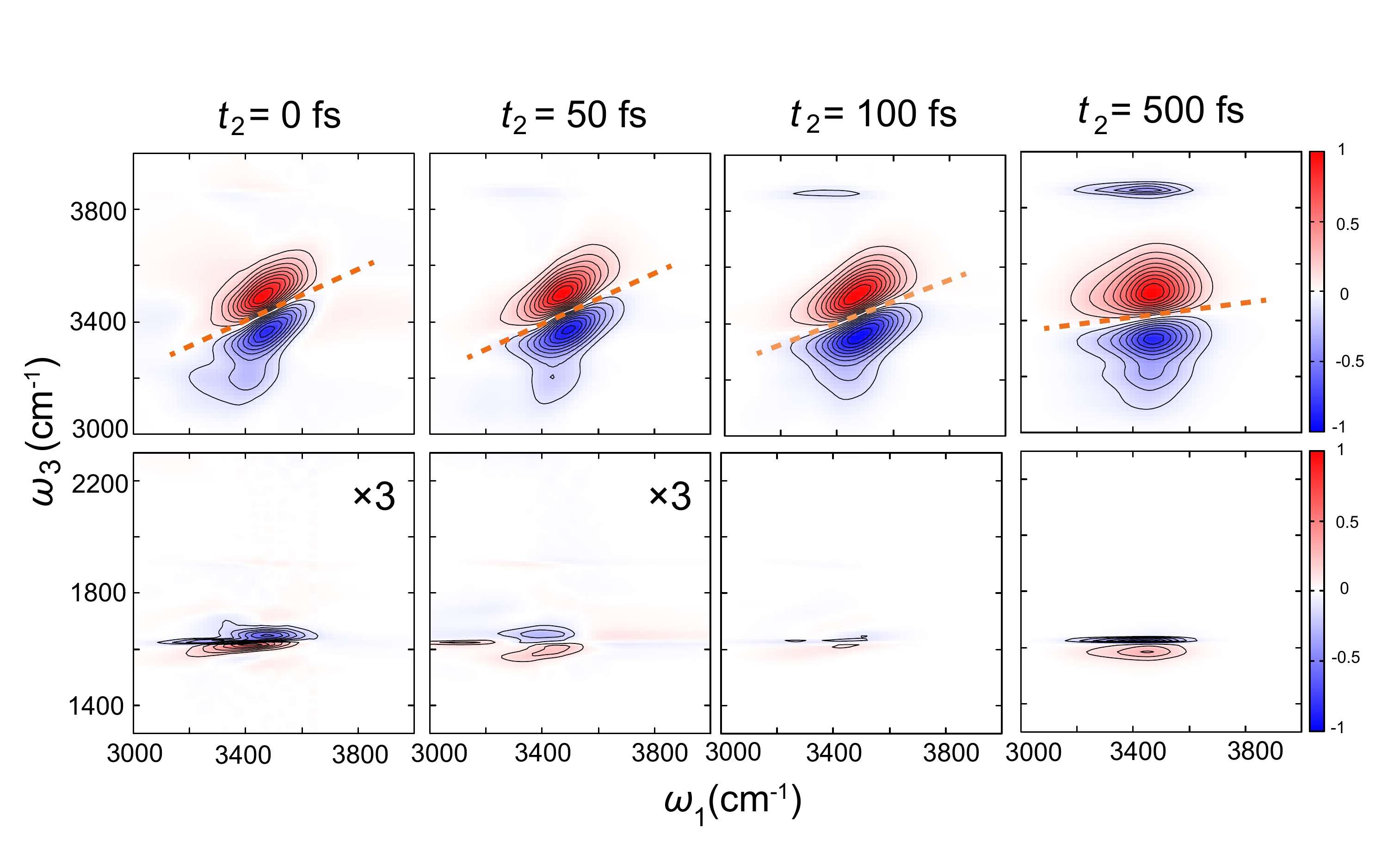}
  \caption{Third-order 2D correlation IR spectra for stretching motion (upper panel) and stretching $\rightarrow$ bending motion (lower panel) calculated for a stronger cubic coupling ($\tilde{g}_{s{s'}^2}=0.5$) than the case in Fig. \ref{fgr:2DIRA}. All spectral intensities are normalized with respect to the absolute value of the maximum peak intensity of each diagonal peak. The direction of the nodal lines (orange dashed lines) in the upper panel represents the extent of correlation between the vibrational coherences of the $t_1$ and $t_3$ periods. The data of off-diagonal peaks are multiplied by 3 for clarity in the cases of $t_2=0$ fs and $50$ fs. }
  \label{fgr:2DIRA2}
\end{figure*}
\begin{figure*}[!t]
  \centering
  \includegraphics[scale=0.6]{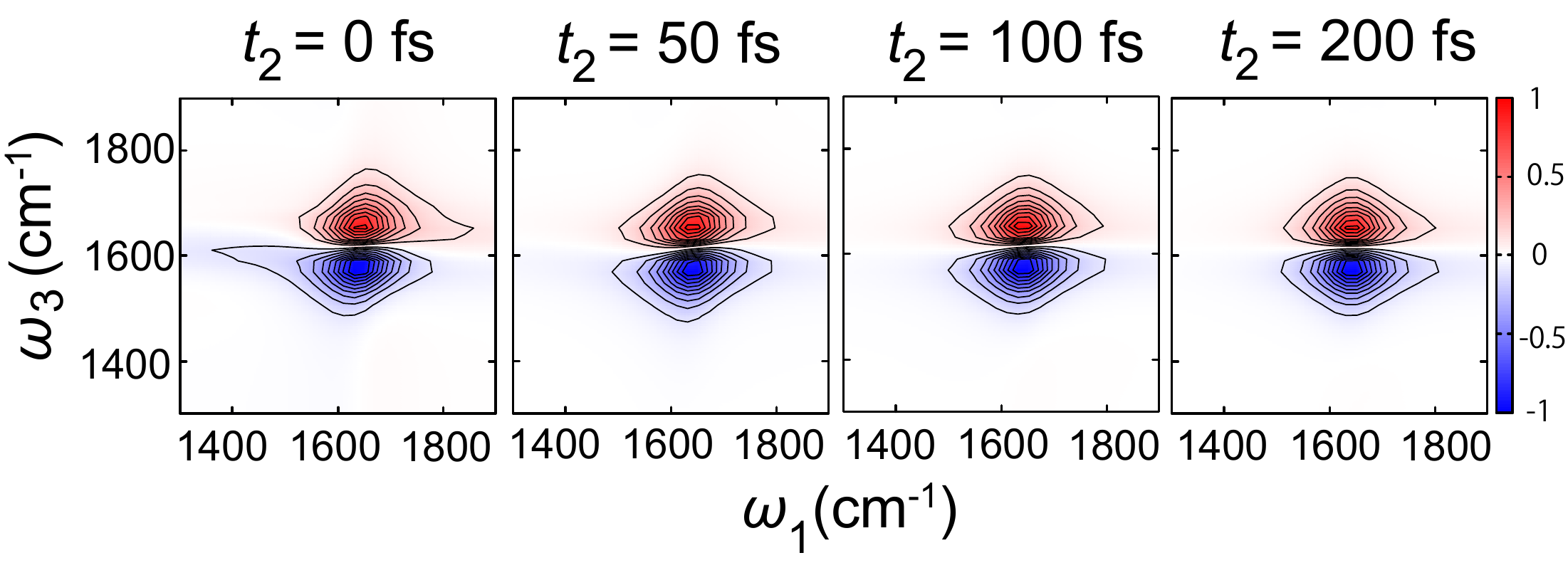}
  \caption{Third-order 2D correlation IR spectra of the bending mode ($s=2$) calculated with the stretching--bending coupling for different $t_2$. All spectral intensities are normalized with respect to the absolute value of the maximum peak intensity in the $t_2=0$ case. }
 \label{fgr:2DIRB2}
\end{figure*}

\section*{Acknowledgments}
The authors are thankful to Professor Shinji Saito and Professor Keisuke Tominaga for helpful discussions. 
Y.T. was supported by JSPS KAKENHI (Grant No. B21H01884). H.T. was supported by JST SPRING (Grant No. JPMJSP2110).

\subsection*{Conflict of Interest}
The authors have no conflicts to disclose.

\section*{Data availability}
The data that support the findings of this study are available from the corresponding author upon reasonable request.

\appendix
\section{2D IR spectra of stretching--bending modes in the enhanced mode-mode coupling case}
\label{sec:2DIRA}
In the investigation of the stretching mode, it was observed experimentally that as $t_2$ increased, the negative (blue) 0--1--2 peak extended in the small $\omega_3$ direction, and a peak, referred to as the hot ground state, appeared above the positive (red) 0--1--0 peak. \cite{ElsaesserDwaynePNAS2008,Tokmakoff2016H2O,VothTokmakoff_St-BendJCP2017,Tokmakoff2022} 
To predict such phenomena with the current approach, one must include various hypothetical factors that are considered important in determining the vibrational process of water, with reference to the 2D spectral profile.
Here, we examine the effects of the cubic coupling strength. Thus we made $\tilde{g}_{12^2}$ 2.5 times larger than the case in Fig. \ref{fgr:2DIRA}, leaving the other parameter values unchanged. Figure \ref{fgr:2DIRA2} illustrates the calculated results. Although less pronounced than the experimentally observed spectrum, the blue 0--1--2 peak extends toward small $\omega_3$ direction from $t_2=0$, which indicates that this peak is instantaneously generated by the coherent transfer rather than a population transfer between the two modes. A small negative peak appears for $t_2\ge 100$fs above the red 0--1--0 peak, although its peak position is much higher than the experimentally observed hot grand state peak. The appearance of this peak indicates that this peak arises from a population transfer rather than coherent transfer as the stretching$\rightarrow$bending peak for $t_2\ge 100$fs in the lower panels. The vibrational coherence, represented by the angle of the orange nodal line, is slightly weaker than in the case in Fig. \ref{fgr:2DIRA}.

\section{2D IR spectra of bending mode}
\label{sec:2DIRB}
In this Appendix, we present third-order 2D IR spectra of the bending mode calculated for the bending$\rightarrow$librational (2$\rightarrow$3) transition case under the same conditions as in Fig. \ref{fgr:bendlib}. This spectrum has been measured experimentally.\cite{Kuroda_BendPCCP2014} In this case, we observe horizontally elongated positive (red) and negative(blue) peaks along $\omega_3=1650$ cm$^{-1}$. The nodal line is always parallel to the $\omega_1$ axis because the noise correlation time is very short ($1/\gamma_2\approx 40$ fs). Note that we have also computed 2D spectra for the case of single bending mode (not shown) and we found that the inclusion of the bending$\rightarrow$librational (2$\rightarrow$3) coupling does not change the 2D profile much. This indicates the effect of population and coherent transfer for the bending$\rightarrow$librational transition is minor. As a result, the peak profile changes only slightly as $t_2$ increases.

\section*{references}
\bibliography{tanimura_publist,TT23}

\begin{thebibliography}{83}%
\makeatletter
\providecommand \@ifxundefined [1]{%
 \@ifx{#1\undefined}
}%
\providecommand \@ifnum [1]{%
 \ifnum #1\expandafter \@firstoftwo
 \else \expandafter \@secondoftwo
 \fi
}%
\providecommand \@ifx [1]{%
 \ifx #1\expandafter \@firstoftwo
 \else \expandafter \@secondoftwo
 \fi
}%
\providecommand \natexlab [1]{#1}%
\providecommand \enquote  [1]{``#1''}%
\providecommand \bibnamefont  [1]{#1}%
\providecommand \bibfnamefont [1]{#1}%
\providecommand \citenamefont [1]{#1}%
\providecommand \href@noop [0]{\@secondoftwo}%
\providecommand \href [0]{\begingroup \@sanitize@url \@href}%
\providecommand \@href[1]{\@@startlink{#1}\@@href}%
\providecommand \@@href[1]{\endgroup#1\@@endlink}%
\providecommand \@sanitize@url [0]{\catcode `\\12\catcode `\$12\catcode
  `\&12\catcode `\#12\catcode `\^12\catcode `\_12\catcode `\%12\relax}%
\providecommand \@@startlink[1]{}%
\providecommand \@@endlink[0]{}%
\providecommand \url  [0]{\begingroup\@sanitize@url \@url }%
\providecommand \@url [1]{\endgroup\@href {#1}{\urlprefix }}%
\providecommand \urlprefix  [0]{URL }%
\providecommand \Eprint [0]{\href }%
\providecommand \doibase [0]{https://doi.org/}%
\providecommand \selectlanguage [0]{\@gobble}%
\providecommand \bibinfo  [0]{\@secondoftwo}%
\providecommand \bibfield  [0]{\@secondoftwo}%
\providecommand \translation [1]{[#1]}%
\providecommand \BibitemOpen [0]{}%
\providecommand \bibitemStop [0]{}%
\providecommand \bibitemNoStop [0]{.\EOS\space}%
\providecommand \EOS [0]{\spacefactor3000\relax}%
\providecommand \BibitemShut  [1]{\csname bibitem#1\endcsname}%
\let\auto@bib@innerbib\@empty
\bibitem [{\citenamefont {Ohmine}\ and\ \citenamefont
  {Saito}(1999)}]{OCSACR1999}%
  \BibitemOpen
  \bibfield  {author} {\bibinfo {author} {\bibfnamefont {I.}~\bibnamefont
  {Ohmine}}\ and\ \bibinfo {author} {\bibfnamefont {S.}~\bibnamefont {Saito}},\
  }\bibfield  {title} {\enquote {\bibinfo {title} {Water dynamics {{:}}
  fluctuation, relaxation, and chemical reactions in hydrogen bond network
  rearrangement},}\ }\href {https://doi.org/10.1021/ar970161g} {\bibfield
  {journal} {\bibinfo  {journal} {Accounts of Chemical Research}\ }\textbf
  {\bibinfo {volume} {32}},\ \bibinfo {pages} {741--749} (\bibinfo {year}
  {1999})}\BibitemShut {NoStop}%
\bibitem [{\citenamefont {Nibbering}\ and\ \citenamefont
  {Elsaesser}(2004)}]{Nibbering2004UltrafastVD}%
  \BibitemOpen
  \bibfield  {author} {\bibinfo {author} {\bibfnamefont {E.~T.~J.}\
  \bibnamefont {Nibbering}}\ and\ \bibinfo {author} {\bibfnamefont
  {T.}~\bibnamefont {Elsaesser}},\ }\bibfield  {title} {\enquote {\bibinfo
  {title} {Ultrafast vibrational dynamics of hydrogen bonds in the condensed
  phase.}}\ }\href {https://doi.org/10.1021/cr020694p} {\bibfield  {journal}
  {\bibinfo  {journal} {Chemical reviews}\ }\textbf {\bibinfo {volume} {104
  4}},\ \bibinfo {pages} {1887--1914} (\bibinfo {year} {2004})}\BibitemShut
  {NoStop}%
\bibitem [{\citenamefont {Steinel}\ \emph {et~al.}(2004)\citenamefont
  {Steinel}, \citenamefont {Asbury}, \citenamefont {Corcelli}, \citenamefont
  {Lawrence}, \citenamefont {Skinner},\ and\ \citenamefont
  {Fayer}}]{SkinnerCPL2004}%
  \BibitemOpen
  \bibfield  {author} {\bibinfo {author} {\bibfnamefont {T.}~\bibnamefont
  {Steinel}}, \bibinfo {author} {\bibfnamefont {J.~B.}\ \bibnamefont {Asbury}},
  \bibinfo {author} {\bibfnamefont {S.}~\bibnamefont {Corcelli}}, \bibinfo
  {author} {\bibfnamefont {C.}~\bibnamefont {Lawrence}}, \bibinfo {author}
  {\bibfnamefont {J.}~\bibnamefont {Skinner}},\ and\ \bibinfo {author}
  {\bibfnamefont {M.}~\bibnamefont {Fayer}},\ }\bibfield  {title} {\enquote
  {\bibinfo {title} {Water dynamics: dependence on local structure probed with
  vibrational echo correlation spectroscopy},}\ }\href
  {https://doi.org/https://doi.org/10.1016/j.cplett.2004.01.042} {\bibfield
  {journal} {\bibinfo  {journal} {Chemical Physics Letters}\ }\textbf {\bibinfo
  {volume} {386}},\ \bibinfo {pages} {295--300} (\bibinfo {year}
  {2004})}\BibitemShut {NoStop}%
\bibitem [{\citenamefont {Bakker}\ and\ \citenamefont
  {Skinner}(2010)}]{BakkerSkinner2010}%
  \BibitemOpen
  \bibfield  {author} {\bibinfo {author} {\bibfnamefont {H.~J.}\ \bibnamefont
  {Bakker}}\ and\ \bibinfo {author} {\bibfnamefont {J.~L.}\ \bibnamefont
  {Skinner}},\ }\bibfield  {title} {\enquote {\bibinfo {title} {Vibrational
  spectroscopy as a probe of structure and dynamics in liquid water},}\ }\href
  {https://doi.org/10.1021/cr9001879} {\bibfield  {journal} {\bibinfo
  {journal} {Chemical Reviews}\ }\textbf {\bibinfo {volume} {110}},\ \bibinfo
  {pages} {1498--1517} (\bibinfo {year} {2010})}\BibitemShut {NoStop}%
\bibitem [{\citenamefont {Fecko}\ \emph {et~al.}(2003)\citenamefont {Fecko},
  \citenamefont {Eaves}, \citenamefont {Loparo}, \citenamefont {Tokmakoff},\
  and\ \citenamefont {Geissler}}]{Tokmakoff2003H2O}%
  \BibitemOpen
  \bibfield  {author} {\bibinfo {author} {\bibfnamefont {C.}~\bibnamefont
  {Fecko}}, \bibinfo {author} {\bibfnamefont {J.}~\bibnamefont {Eaves}},
  \bibinfo {author} {\bibfnamefont {J.}~\bibnamefont {Loparo}}, \bibinfo
  {author} {\bibfnamefont {A.}~\bibnamefont {Tokmakoff}},\ and\ \bibinfo
  {author} {\bibfnamefont {P.}~\bibnamefont {Geissler}},\ }\bibfield  {title}
  {\enquote {\bibinfo {title} {Ultrafast hydrogen-bond dynamics in the infrared
  spectroscopy of water},}\ }\href {https://doi.org/10.1126/science.1087251}
  {\bibfield  {journal} {\bibinfo  {journal} {SCIENCE}\ }\textbf {\bibinfo
  {volume} {301}},\ \bibinfo {pages} {1698--1702} (\bibinfo {year}
  {2003})}\BibitemShut {NoStop}%
\bibitem [{\citenamefont {Rey}, \citenamefont {Møller},\ and\ \citenamefont
  {Hynes}(2004)}]{Hynes2004}%
  \BibitemOpen
  \bibfield  {author} {\bibinfo {author} {\bibfnamefont {R.}~\bibnamefont
  {Rey}}, \bibinfo {author} {\bibfnamefont {K.~B.}\ \bibnamefont {Møller}},\
  and\ \bibinfo {author} {\bibfnamefont {J.~T.}\ \bibnamefont {Hynes}},\
  }\bibfield  {title} {\enquote {\bibinfo {title} {Ultrafast vibrational
  population dynamics of water and related systems, a theoretical
  perspective},}\ }\href {https://doi.org/10.1021/cr020675f} {\bibfield
  {journal} {\bibinfo  {journal} {Chemical Reviews}\ }\textbf {\bibinfo
  {volume} {104}},\ \bibinfo {pages} {1915--1928} (\bibinfo {year}
  {2004})}\BibitemShut {NoStop}%
\bibitem [{\citenamefont {Liu}\ \emph {et~al.}(2011)\citenamefont {Liu},
  \citenamefont {Miller}, \citenamefont {Fanourgakis}, \citenamefont
  {Xantheas}, \citenamefont {Imoto},\ and\ \citenamefont
  {Saito}}]{Imoto_JCP135}%
  \BibitemOpen
  \bibfield  {author} {\bibinfo {author} {\bibfnamefont {J.}~\bibnamefont
  {Liu}}, \bibinfo {author} {\bibfnamefont {W.~H.}\ \bibnamefont {Miller}},
  \bibinfo {author} {\bibfnamefont {G.~S.}\ \bibnamefont {Fanourgakis}},
  \bibinfo {author} {\bibfnamefont {S.~S.}\ \bibnamefont {Xantheas}}, \bibinfo
  {author} {\bibfnamefont {S.}~\bibnamefont {Imoto}},\ and\ \bibinfo {author}
  {\bibfnamefont {S.}~\bibnamefont {Saito}},\ }\bibfield  {title} {\enquote
  {\bibinfo {title} {Insights in quantum dynamical effects in the infrared
  spectroscopy of liquid water from a semiclassical study with an ab
  initio-based flexible and polarizable force field},}\ }\href
  {https://doi.org/10.1063/1.3670960} {\bibfield  {journal} {\bibinfo
  {journal} {The Journal of Chemical Physics}\ }\textbf {\bibinfo {volume}
  {135}},\ \bibinfo {pages} {244503} (\bibinfo {year} {2011})},\ \Eprint
  {https://arxiv.org/abs/https://doi.org/10.1063/1.3670960}
  {https://doi.org/10.1063/1.3670960} \BibitemShut {NoStop}%
\bibitem [{\citenamefont {Perakis}\ \emph {et~al.}(2016)\citenamefont
  {Perakis}, \citenamefont {De~Marco}, \citenamefont {Shalit}, \citenamefont
  {Tang}, \citenamefont {Kann}, \citenamefont {Kühne}, \citenamefont {Torre},
  \citenamefont {Bonn},\ and\ \citenamefont {Nagata}}]{NagataChemRev2016}%
  \BibitemOpen
  \bibfield  {author} {\bibinfo {author} {\bibfnamefont {F.}~\bibnamefont
  {Perakis}}, \bibinfo {author} {\bibfnamefont {L.}~\bibnamefont {De~Marco}},
  \bibinfo {author} {\bibfnamefont {A.}~\bibnamefont {Shalit}}, \bibinfo
  {author} {\bibfnamefont {F.}~\bibnamefont {Tang}}, \bibinfo {author}
  {\bibfnamefont {Z.~R.}\ \bibnamefont {Kann}}, \bibinfo {author}
  {\bibfnamefont {T.~D.}\ \bibnamefont {Kühne}}, \bibinfo {author}
  {\bibfnamefont {R.}~\bibnamefont {Torre}}, \bibinfo {author} {\bibfnamefont
  {M.}~\bibnamefont {Bonn}},\ and\ \bibinfo {author} {\bibfnamefont
  {Y.}~\bibnamefont {Nagata}},\ }\bibfield  {title} {\enquote {\bibinfo {title}
  {Vibrational spectroscopy and dynamics of water},}\ }\href
  {https://doi.org/10.1021/acs.chemrev.5b00640} {\bibfield  {journal} {\bibinfo
   {journal} {Chemical Reviews}\ }\textbf {\bibinfo {volume} {116}},\ \bibinfo
  {pages} {7590--7607} (\bibinfo {year} {2016})},\ \Eprint
  {https://arxiv.org/abs/https://doi.org/10.1021/acs.chemrev.5b00640}
  {https://doi.org/10.1021/acs.chemrev.5b00640} \BibitemShut {NoStop}%
\bibitem [{\citenamefont {Ohmine}\ and\ \citenamefont
  {Tanaka}(1993)}]{Ohmine_ChemRev93}%
  \BibitemOpen
  \bibfield  {author} {\bibinfo {author} {\bibfnamefont {I.}~\bibnamefont
  {Ohmine}}\ and\ \bibinfo {author} {\bibfnamefont {H.}~\bibnamefont
  {Tanaka}},\ }\bibfield  {title} {\enquote {\bibinfo {title} {Fluctuation,
  relaxations, and hydration in liquid water. hydrogen-bond rearrangement
  dynamics},}\ }\href {https://doi.org/10.1021/cr00023a011} {\bibfield
  {journal} {\bibinfo  {journal} {Chemical Reviews}\ }\textbf {\bibinfo
  {volume} {93}},\ \bibinfo {pages} {2545--2566} (\bibinfo {year} {1993})},\
  \Eprint {https://arxiv.org/abs/https://doi.org/10.1021/cr00023a011}
  {https://doi.org/10.1021/cr00023a011} \BibitemShut {NoStop}%
\bibitem [{\citenamefont {Yagasaki}\ and\ \citenamefont
  {Saito}(2009)}]{Yagasaki_ACR42}%
  \BibitemOpen
  \bibfield  {author} {\bibinfo {author} {\bibfnamefont {T.}~\bibnamefont
  {Yagasaki}}\ and\ \bibinfo {author} {\bibfnamefont {S.}~\bibnamefont
  {Saito}},\ }\bibfield  {title} {\enquote {\bibinfo {title} {Molecular
  dynamics simulation of nonlinear spectroscopies of intermolecular motions in
  liquid water},}\ }\href {https://doi.org/10.1021/ar900007s} {\bibfield
  {journal} {\bibinfo  {journal} {Accounts of Chemical Research}\ }\textbf
  {\bibinfo {volume} {42}},\ \bibinfo {pages} {1250--1258} (\bibinfo {year}
  {2009})},\ \Eprint {https://arxiv.org/abs/https://doi.org/10.1021/ar900007s}
  {https://doi.org/10.1021/ar900007s} \BibitemShut {NoStop}%
\bibitem [{\citenamefont {Yagasaki}\ and\ \citenamefont
  {Saito}(2013)}]{Yagasaki_ARPC64}%
  \BibitemOpen
  \bibfield  {author} {\bibinfo {author} {\bibfnamefont {T.}~\bibnamefont
  {Yagasaki}}\ and\ \bibinfo {author} {\bibfnamefont {S.}~\bibnamefont
  {Saito}},\ }\bibfield  {title} {\enquote {\bibinfo {title} {Fluctuations and
  relaxation dynamics of liquid water revealed by linear and nonlinear
  spectroscopy},}\ }\href
  {https://doi.org/10.1146/annurev-physchem-040412-110150} {\bibfield
  {journal} {\bibinfo  {journal} {Annual Review of Physical Chemistry}\
  }\textbf {\bibinfo {volume} {64}},\ \bibinfo {pages} {55--75} (\bibinfo
  {year} {2013})},\ \Eprint
  {https://arxiv.org/abs/https://doi.org/10.1146/annurev-physchem-040412-110150}
  {https://doi.org/10.1146/annurev-physchem-040412-110150} \BibitemShut
  {NoStop}%
\bibitem [{\citenamefont {Imoto}, \citenamefont {Xantheas},\ and\ \citenamefont
  {Saito}(2015)}]{Imotobend-lib2015}%
  \BibitemOpen
  \bibfield  {author} {\bibinfo {author} {\bibfnamefont {S.}~\bibnamefont
  {Imoto}}, \bibinfo {author} {\bibfnamefont {S.~S.}\ \bibnamefont
  {Xantheas}},\ and\ \bibinfo {author} {\bibfnamefont {S.}~\bibnamefont
  {Saito}},\ }\bibfield  {title} {\enquote {\bibinfo {title} {Ultrafast
  dynamics of liquid water: Energy relaxation and transfer processes of the
  \uppercase{OH} stretch and the \uppercase{HOH} bend},}\ }\href
  {https://doi.org/10.1021/acs.jpcb.5b02589} {\bibfield  {journal} {\bibinfo
  {journal} {The Journal of Physical Chemistry B}\ }\textbf {\bibinfo {volume}
  {119}},\ \bibinfo {pages} {11068--11078} (\bibinfo {year} {2015})},\ \bibinfo
  {note} {pMID: 26042611},\ \Eprint
  {https://arxiv.org/abs/https://doi.org/10.1021/acs.jpcb.5b02589}
  {https://doi.org/10.1021/acs.jpcb.5b02589} \BibitemShut {NoStop}%
\bibitem [{\citenamefont {Huse}\ \emph {et~al.}(2005)\citenamefont {Huse},
  \citenamefont {Ashihara}, \citenamefont {Nibbering},\ and\ \citenamefont
  {Elsaesser}}]{ElsaesserCPL2005}%
  \BibitemOpen
  \bibfield  {author} {\bibinfo {author} {\bibfnamefont {N.}~\bibnamefont
  {Huse}}, \bibinfo {author} {\bibfnamefont {S.}~\bibnamefont {Ashihara}},
  \bibinfo {author} {\bibfnamefont {E.}~\bibnamefont {Nibbering}},\ and\
  \bibinfo {author} {\bibfnamefont {T.}~\bibnamefont {Elsaesser}},\ }\bibfield
  {title} {\enquote {\bibinfo {title} {Ultrafast vibrational relaxation of
  \uppercase{O}-\uppercase{H} bending and librational excitations in liquid
  \uppercase{H}$_2$\uppercase{O}},}\ }\href
  {https://doi.org/10.1016/j.cplett.2005.02.007} {\bibfield  {journal}
  {\bibinfo  {journal} {CHEMICAL PHYSICS LETTERS}\ }\textbf {\bibinfo {volume}
  {404}},\ \bibinfo {pages} {389--393} (\bibinfo {year} {2005})}\BibitemShut
  {NoStop}%
\bibitem [{\citenamefont {Ramasesha}\ \emph {et~al.}(2013)\citenamefont
  {Ramasesha}, \citenamefont {De~Marco}, \citenamefont {Mandal},\ and\
  \citenamefont {Tokmakoff}}]{TokmakoffNat2013}%
  \BibitemOpen
  \bibfield  {author} {\bibinfo {author} {\bibfnamefont {K.}~\bibnamefont
  {Ramasesha}}, \bibinfo {author} {\bibfnamefont {L.}~\bibnamefont {De~Marco}},
  \bibinfo {author} {\bibfnamefont {A.}~\bibnamefont {Mandal}},\ and\ \bibinfo
  {author} {\bibfnamefont {A.}~\bibnamefont {Tokmakoff}},\ }\bibfield  {title}
  {\enquote {\bibinfo {title} {Water vibrations have strongly mixed intra- and
  intermolecular character},}\ }\href {https://doi.org/10.1038/nchem.1757}
  {\bibfield  {journal} {\bibinfo  {journal} {Nature Chemistry}\ }\textbf
  {\bibinfo {volume} {5}},\ \bibinfo {pages} {935--940} (\bibinfo {year}
  {2013})}\BibitemShut {NoStop}%
\bibitem [{\citenamefont {Bertie}\ and\ \citenamefont
  {Lan}(1996)}]{Bertie96IRexp}%
  \BibitemOpen
  \bibfield  {author} {\bibinfo {author} {\bibfnamefont {J.~E.}\ \bibnamefont
  {Bertie}}\ and\ \bibinfo {author} {\bibfnamefont {Z.}~\bibnamefont {Lan}},\
  }\bibfield  {title} {\enquote {\bibinfo {title} {Infrared intensities of
  liquids xx: The intensity of the \uppercase{OH} stretching band of liquid
  water revisited, and the best current values of the optical constants of
  \uppercase{H}$_2$\uppercase{O}(l) at 25{\textdegree}c between 15,000 and 1
  cm-1},}\ }\href {https://opg.optica.org/as/abstract.cfm?URI=as-50-8-1047}
  {\bibfield  {journal} {\bibinfo  {journal} {Appl. Spectrosc.}\ }\textbf
  {\bibinfo {volume} {50}},\ \bibinfo {pages} {1047--1057} (\bibinfo {year}
  {1996})}\BibitemShut {NoStop}%
\bibitem [{\citenamefont {Maréchal}(2011)}]{IRexp2011}%
  \BibitemOpen
  \bibfield  {author} {\bibinfo {author} {\bibfnamefont {Y.}~\bibnamefont
  {Maréchal}},\ }\bibfield  {title} {\enquote {\bibinfo {title} {The molecular
  structure of liquid water delivered by absorption spectroscopy in the whole
  \uppercase{IR} region completed with thermodynamics data},}\ }\href
  {https://doi.org/https://doi.org/10.1016/j.molstruc.2011.07.054} {\bibfield
  {journal} {\bibinfo  {journal} {Journal of Molecular Structure}\ }\textbf
  {\bibinfo {volume} {1004}},\ \bibinfo {pages} {146--155} (\bibinfo {year}
  {2011})}\BibitemShut {NoStop}%
\bibitem [{\citenamefont {Brooker}\ \emph {et~al.}(1989)\citenamefont
  {Brooker}, \citenamefont {Hancock}, \citenamefont {Rice},\ and\ \citenamefont
  {Shapter}}]{Brooker1989Ramanexp}%
  \BibitemOpen
  \bibfield  {author} {\bibinfo {author} {\bibfnamefont {M.~H.}\ \bibnamefont
  {Brooker}}, \bibinfo {author} {\bibfnamefont {G.~W.}\ \bibnamefont
  {Hancock}}, \bibinfo {author} {\bibfnamefont {B.}~\bibnamefont {Rice}},\ and\
  \bibinfo {author} {\bibfnamefont {J.~G.}\ \bibnamefont {Shapter}},\
  }\bibfield  {title} {\enquote {\bibinfo {title} {\uppercase{R}aman frequency
  and intensity studies of liquid \uppercase{H}$_2$\uppercase{O},
  \uppercase{H}$_2$18\uppercase{O} and \uppercase{D}$_2$\uppercase{O}},}\
  }\href@noop {} {\bibfield  {journal} {\bibinfo  {journal} {Journal of Raman
  Spectroscopy}\ }\textbf {\bibinfo {volume} {20}},\ \bibinfo {pages}
  {683--694} (\bibinfo {year} {1989})}\BibitemShut {NoStop}%
\bibitem [{\citenamefont {Pattenaude}, \citenamefont {Streacker},\ and\
  \citenamefont {Ben-Amotz}(2018)}]{Raman2018exp}%
  \BibitemOpen
  \bibfield  {author} {\bibinfo {author} {\bibfnamefont {S.~R.}\ \bibnamefont
  {Pattenaude}}, \bibinfo {author} {\bibfnamefont {L.~M.}\ \bibnamefont
  {Streacker}},\ and\ \bibinfo {author} {\bibfnamefont {D.}~\bibnamefont
  {Ben-Amotz}},\ }\bibfield  {title} {\enquote {\bibinfo {title} {Temperature
  and polarization dependent \uppercase{R}aman spectra of liquid
  \uppercase{H}$_2$\uppercase{O} and \uppercase{D}$_2$\uppercase{O}},}\ }\href
  {https://doi.org/https://doi.org/10.1002/jrs.5465} {\bibfield  {journal}
  {\bibinfo  {journal} {Journal of Raman Spectroscopy}\ }\textbf {\bibinfo
  {volume} {49}},\ \bibinfo {pages} {1860--1866} (\bibinfo {year}
  {2018})}\BibitemShut {NoStop}%
\bibitem [{\citenamefont {Heisler}, \citenamefont {Mazur},\ and\ \citenamefont
  {Meech}(2017)}]{HEISLER201745}%
  \BibitemOpen
  \bibfield  {author} {\bibinfo {author} {\bibfnamefont {I.~A.}\ \bibnamefont
  {Heisler}}, \bibinfo {author} {\bibfnamefont {K.}~\bibnamefont {Mazur}},\
  and\ \bibinfo {author} {\bibfnamefont {S.~R.}\ \bibnamefont {Meech}},\
  }\bibfield  {title} {\enquote {\bibinfo {title} {\uppercase{R}aman
  vibrational dynamics of hydrated ions in the low-frequency spectral
  region},}\ }\href
  {https://doi.org/https://doi.org/10.1016/j.molliq.2016.09.066} {\bibfield
  {journal} {\bibinfo  {journal} {Journal of Molecular Liquids}\ }\textbf
  {\bibinfo {volume} {228}},\ \bibinfo {pages} {45--53} (\bibinfo {year}
  {2017})},\ \bibinfo {note} {from simple liquids to macromolecular solutions:
  recent experimental and theoretical developments. In Honor of the 70th
  birthday of Vojko Vlachy}\BibitemShut {NoStop}%
\bibitem [{\citenamefont {Mukamel}(1999)}]{mukamel1999principles}%
  \BibitemOpen
  \bibfield  {author} {\bibinfo {author} {\bibfnamefont {S.}~\bibnamefont
  {Mukamel}},\ }\href@noop {} {\emph {\bibinfo {title} {Principles of nonlinear
  optical spectroscopy}}},\ \bibinfo {number} {6}\ (\bibinfo  {publisher}
  {Oxford University Press on Demand},\ \bibinfo {year} {1999})\BibitemShut
  {NoStop}%
\bibitem [{\citenamefont {Tanimura}\ and\ \citenamefont
  {Mukamel}(1993)}]{TM93JCP}%
  \BibitemOpen
  \bibfield  {author} {\bibinfo {author} {\bibfnamefont {Y.}~\bibnamefont
  {Tanimura}}\ and\ \bibinfo {author} {\bibfnamefont {S.}~\bibnamefont
  {Mukamel}},\ }\bibfield  {title} {\enquote {\bibinfo {title} {Two-dimensional
  femtosecond vibrational spectroscopy of liquids},}\ }\href
  {https://doi.org/10.1063/1.465484} {\bibfield  {journal} {\bibinfo  {journal}
  {The Journal of Chemical Physics}\ }\textbf {\bibinfo {volume} {99}},\
  \bibinfo {pages} {9496--9511} (\bibinfo {year} {1993})}\BibitemShut {NoStop}%
\bibitem [{\citenamefont {Tanimura}\ and\ \citenamefont
  {Ishizaki}(2009)}]{TI09ACR}%
  \BibitemOpen
  \bibfield  {author} {\bibinfo {author} {\bibfnamefont {Y.}~\bibnamefont
  {Tanimura}}\ and\ \bibinfo {author} {\bibfnamefont {A.}~\bibnamefont
  {Ishizaki}},\ }\bibfield  {title} {\enquote {\bibinfo {title} {Modeling,
  calculating, and analyzing multidimensional vibrational spectroscopies},}\
  }\href {https://doi.org/10.1021/ar9000444} {\bibfield  {journal} {\bibinfo
  {journal} {Accounts of Chemical Research}\ }\textbf {\bibinfo {volume}
  {42}},\ \bibinfo {pages} {1270--1279} (\bibinfo {year} {2009})}\BibitemShut
  {NoStop}%
\bibitem [{\citenamefont {Cho}(2009)}]{Cho2009}%
  \BibitemOpen
  \bibfield  {author} {\bibinfo {author} {\bibfnamefont {M.}~\bibnamefont
  {Cho}},\ }\href {https://doi.org/https://doi.org/10.1201/9781420084306}
  {\emph {\bibinfo {title} {Two-Dimensional Optical Spectroscopy}}}\ (\bibinfo
  {publisher} {CRC Press},\ \bibinfo {year} {2009})\BibitemShut {NoStop}%
\bibitem [{\citenamefont {Hamm}\ and\ \citenamefont
  {Zanni}(2011)}]{Hamm2011ConceptsAM}%
  \BibitemOpen
  \bibfield  {author} {\bibinfo {author} {\bibfnamefont {P.}~\bibnamefont
  {Hamm}}\ and\ \bibinfo {author} {\bibfnamefont {M.~T.}\ \bibnamefont
  {Zanni}},\ }\href {https://doi.org/https://doi.org/10.1017/CBO9780511675935}
  {\emph {\bibinfo {title} {Concepts and Methods of 2D Infrared
  Spectroscopy}}}\ (\bibinfo  {publisher} {Cambridge University Press},\
  \bibinfo {year} {2011})\BibitemShut {NoStop}%
\bibitem [{\citenamefont {Kraemer}\ \emph {et~al.}(2008)\citenamefont
  {Kraemer}, \citenamefont {Cowan}, \citenamefont {Paarmann}, \citenamefont
  {Huse}, \citenamefont {Nibbering}, \citenamefont {Elsaesser},\ and\
  \citenamefont {Miller}}]{ElsaesserDwaynePNAS2008}%
  \BibitemOpen
  \bibfield  {author} {\bibinfo {author} {\bibfnamefont {D.}~\bibnamefont
  {Kraemer}}, \bibinfo {author} {\bibfnamefont {M.~L.}\ \bibnamefont {Cowan}},
  \bibinfo {author} {\bibfnamefont {A.}~\bibnamefont {Paarmann}}, \bibinfo
  {author} {\bibfnamefont {N.}~\bibnamefont {Huse}}, \bibinfo {author}
  {\bibfnamefont {E.~T.~J.}\ \bibnamefont {Nibbering}}, \bibinfo {author}
  {\bibfnamefont {T.}~\bibnamefont {Elsaesser}},\ and\ \bibinfo {author}
  {\bibfnamefont {R.~J.~D.}\ \bibnamefont {Miller}},\ }\bibfield  {title}
  {\enquote {\bibinfo {title} {Temperature dependence of the two-dimensional
  infrared spectrum of liquid \uppercase{H}$_2$\uppercase{O}},}\ }\href
  {https://doi.org/10.1073/pnas.0705792105} {\bibfield  {journal} {\bibinfo
  {journal} {Proceedings of the National Academy of Sciences}\ }\textbf
  {\bibinfo {volume} {105}},\ \bibinfo {pages} {437--442} (\bibinfo {year}
  {2008})}\BibitemShut {NoStop}%
\bibitem [{\citenamefont {De~Marco}\ \emph {et~al.}(2016)\citenamefont
  {De~Marco}, \citenamefont {Fournier}, \citenamefont {Thämer}, \citenamefont
  {Carpenter},\ and\ \citenamefont {Tokmakoff}}]{Tokmakoff2016H2O}%
  \BibitemOpen
  \bibfield  {author} {\bibinfo {author} {\bibfnamefont {L.}~\bibnamefont
  {De~Marco}}, \bibinfo {author} {\bibfnamefont {J.~A.}\ \bibnamefont
  {Fournier}}, \bibinfo {author} {\bibfnamefont {M.}~\bibnamefont {Thämer}},
  \bibinfo {author} {\bibfnamefont {W.}~\bibnamefont {Carpenter}},\ and\
  \bibinfo {author} {\bibfnamefont {A.}~\bibnamefont {Tokmakoff}},\ }\bibfield
  {title} {\enquote {\bibinfo {title} {Anharmonic exciton dynamics and energy
  dissipation in liquid water from two-dimensional infrared spectroscopy},}\
  }\href {https://doi.org/10.1063/1.4961752} {\bibfield  {journal} {\bibinfo
  {journal} {The Journal of Chemical Physics}\ }\textbf {\bibinfo {volume}
  {145}},\ \bibinfo {pages} {094501} (\bibinfo {year} {2016})},\ \Eprint
  {https://arxiv.org/abs/https://aip.scitation.org/doi/pdf/10.1063/1.4961752}
  {https://aip.scitation.org/doi/pdf/10.1063/1.4961752} \BibitemShut {NoStop}%
\bibitem [{\citenamefont {Carpenter}\ \emph {et~al.}(2017)\citenamefont
  {Carpenter}, \citenamefont {Fournier}, \citenamefont {Biswas}, \citenamefont
  {Voth},\ and\ \citenamefont {Tokmakoff}}]{VothTokmakoff_St-BendJCP2017}%
  \BibitemOpen
  \bibfield  {author} {\bibinfo {author} {\bibfnamefont {W.~B.}\ \bibnamefont
  {Carpenter}}, \bibinfo {author} {\bibfnamefont {J.~A.}\ \bibnamefont
  {Fournier}}, \bibinfo {author} {\bibfnamefont {R.}~\bibnamefont {Biswas}},
  \bibinfo {author} {\bibfnamefont {G.~A.}\ \bibnamefont {Voth}},\ and\
  \bibinfo {author} {\bibfnamefont {A.}~\bibnamefont {Tokmakoff}},\ }\bibfield
  {title} {\enquote {\bibinfo {title} {Delocalization and stretch-bend mixing
  of the \uppercase{HOH} bend in liquid water},}\ }\href
  {https://doi.org/10.1063/1.4987153} {\bibfield  {journal} {\bibinfo
  {journal} {The Journal of Chemical Physics}\ }\textbf {\bibinfo {volume}
  {147}},\ \bibinfo {pages} {084503} (\bibinfo {year} {2017})},\ \Eprint
  {https://arxiv.org/abs/https://doi.org/10.1063/1.4987153}
  {https://doi.org/10.1063/1.4987153} \BibitemShut {NoStop}%
\bibitem [{\citenamefont {Lewis}\ \emph {et~al.}(2022)\citenamefont {Lewis},
  \citenamefont {Dereka}, \citenamefont {Zhang}, \citenamefont {Maginn},\ and\
  \citenamefont {Tokmakoff}}]{Tokmakoff2022}%
  \BibitemOpen
  \bibfield  {author} {\bibinfo {author} {\bibfnamefont {N.~H.~C.}\
  \bibnamefont {Lewis}}, \bibinfo {author} {\bibfnamefont {B.}~\bibnamefont
  {Dereka}}, \bibinfo {author} {\bibfnamefont {Y.}~\bibnamefont {Zhang}},
  \bibinfo {author} {\bibfnamefont {E.~J.}\ \bibnamefont {Maginn}},\ and\
  \bibinfo {author} {\bibfnamefont {A.}~\bibnamefont {Tokmakoff}},\ }\bibfield
  {title} {\enquote {\bibinfo {title} {From networked to isolated: Observing
  water hydrogen bonds in concentrated electrolytes with two-dimensional
  infrared spectroscopy},}\ }\href {https://doi.org/10.1021/acs.jpcb.2c03341}
  {\bibfield  {journal} {\bibinfo  {journal} {The Journal of Physical Chemistry
  B}\ }\textbf {\bibinfo {volume} {126}},\ \bibinfo {pages} {5305--5319}
  (\bibinfo {year} {2022})},\ \bibinfo {note} {pMID: 35829623},\ \Eprint
  {https://arxiv.org/abs/https://doi.org/10.1021/acs.jpcb.2c03341}
  {https://doi.org/10.1021/acs.jpcb.2c03341} \BibitemShut {NoStop}%
\bibitem [{\citenamefont {Chuntonov}, \citenamefont {Kumar},\ and\
  \citenamefont {Kuroda}(2014)}]{Kuroda_BendPCCP2014}%
  \BibitemOpen
  \bibfield  {author} {\bibinfo {author} {\bibfnamefont {L.}~\bibnamefont
  {Chuntonov}}, \bibinfo {author} {\bibfnamefont {R.}~\bibnamefont {Kumar}},\
  and\ \bibinfo {author} {\bibfnamefont {D.~G.}\ \bibnamefont {Kuroda}},\
  }\bibfield  {title} {\enquote {\bibinfo {title} {Non-linear infrared
  spectroscopy of the water bending mode: direct experimental evidence of
  hydration shell reorganization?}}\ }\href
  {https://doi.org/10.1039/C4CP00643G} {\bibfield  {journal} {\bibinfo
  {journal} {Phys. Chem. Chem. Phys.}\ }\textbf {\bibinfo {volume} {16}},\
  \bibinfo {pages} {13172--13181} (\bibinfo {year} {2014})}\BibitemShut
  {NoStop}%
\bibitem [{\citenamefont {Hamm}\ and\ \citenamefont
  {Savolainen}(2012)}]{HammTHz2012}%
  \BibitemOpen
  \bibfield  {author} {\bibinfo {author} {\bibfnamefont {P.}~\bibnamefont
  {Hamm}}\ and\ \bibinfo {author} {\bibfnamefont {J.}~\bibnamefont
  {Savolainen}},\ }\bibfield  {title} {\enquote {\bibinfo {title}
  {Two-dimensional-\uppercase{R}aman-\uppercase{T}erahertz spectroscopy of
  water: Theory},}\ }\href {https://doi.org/10.1063/1.3691601} {\bibfield
  {journal} {\bibinfo  {journal} {The Journal of Chemical Physics}\ }\textbf
  {\bibinfo {volume} {136}},\ \bibinfo {pages} {094516} (\bibinfo {year}
  {2012})},\ \Eprint {https://arxiv.org/abs/https://doi.org/10.1063/1.3691601}
  {https://doi.org/10.1063/1.3691601} \BibitemShut {NoStop}%
\bibitem [{\citenamefont {Savolainen}, \citenamefont {Ahmed},\ and\
  \citenamefont {Hamm}(2013)}]{Hamm2013PNAS}%
  \BibitemOpen
  \bibfield  {author} {\bibinfo {author} {\bibfnamefont {J.}~\bibnamefont
  {Savolainen}}, \bibinfo {author} {\bibfnamefont {S.}~\bibnamefont {Ahmed}},\
  and\ \bibinfo {author} {\bibfnamefont {P.}~\bibnamefont {Hamm}},\ }\bibfield
  {title} {\enquote {\bibinfo {title} {Two-dimensional
  \uppercase{R}aman--\uppercase{T}erahertz spectroscopy of water},}\ }\href
  {https://doi.org/10.1073/pnas.1317459110} {\bibfield  {journal} {\bibinfo
  {journal} {Proceedings of the National Academy of Sciences}\ }\textbf
  {\bibinfo {volume} {110}},\ \bibinfo {pages} {20402--20407} (\bibinfo {year}
  {2013})},\ \Eprint
  {https://arxiv.org/abs/https://www.pnas.org/doi/pdf/10.1073/pnas.1317459110}
  {https://www.pnas.org/doi/pdf/10.1073/pnas.1317459110} \BibitemShut {NoStop}%
\bibitem [{\citenamefont {Hamm}(2014)}]{hamm2014}%
  \BibitemOpen
  \bibfield  {author} {\bibinfo {author} {\bibfnamefont {P.}~\bibnamefont
  {Hamm}},\ }\bibfield  {title} {\enquote {\bibinfo {title}
  {{{2D-\uppercase{R}aman-\uppercase{TH}z}} spectroscopy: {{A}} sensitive test
  of polarizable water models},}\ }\href {https://doi.org/10.1063/1.4901216}
  {\bibfield  {journal} {\bibinfo  {journal} {The Journal of Chemical Physics}\
  }\textbf {\bibinfo {volume} {141}},\ \bibinfo {pages} {184201} (\bibinfo
  {year} {2014})}\BibitemShut {NoStop}%
\bibitem [{\citenamefont {Hamm}\ and\ \citenamefont
  {Shalit}(2017)}]{HammPerspH2O2017}%
  \BibitemOpen
  \bibfield  {author} {\bibinfo {author} {\bibfnamefont {P.}~\bibnamefont
  {Hamm}}\ and\ \bibinfo {author} {\bibfnamefont {A.}~\bibnamefont {Shalit}},\
  }\bibfield  {title} {\enquote {\bibinfo {title} {Perspective: Echoes in
  2d-\uppercase{R}aman-\uppercase{TH}z spectroscopy},}\ }\href
  {https://doi.org/10.1063/1.4979288} {\bibfield  {journal} {\bibinfo
  {journal} {The Journal of Chemical Physics}\ }\textbf {\bibinfo {volume}
  {146}},\ \bibinfo {pages} {130901} (\bibinfo {year} {2017})},\ \Eprint
  {https://arxiv.org/abs/https://doi.org/10.1063/1.4979288}
  {https://doi.org/10.1063/1.4979288} \BibitemShut {NoStop}%
\bibitem [{\citenamefont {Grechko}\ \emph {et~al.}(2018)\citenamefont
  {Grechko}, \citenamefont {Hasegawa}, \citenamefont {D'Angelo}, \citenamefont
  {Ito}, \citenamefont {Turchinovich}, \citenamefont {Nagata},\ and\
  \citenamefont {Bonn}}]{grechko2018}%
  \BibitemOpen
  \bibfield  {author} {\bibinfo {author} {\bibfnamefont {M.}~\bibnamefont
  {Grechko}}, \bibinfo {author} {\bibfnamefont {T.}~\bibnamefont {Hasegawa}},
  \bibinfo {author} {\bibfnamefont {F.}~\bibnamefont {D'Angelo}}, \bibinfo
  {author} {\bibfnamefont {H.}~\bibnamefont {Ito}}, \bibinfo {author}
  {\bibfnamefont {D.}~\bibnamefont {Turchinovich}}, \bibinfo {author}
  {\bibfnamefont {Y.}~\bibnamefont {Nagata}},\ and\ \bibinfo {author}
  {\bibfnamefont {M.}~\bibnamefont {Bonn}},\ }\bibfield  {title} {\enquote
  {\bibinfo {title} {Coupling between intra- and intermolecular motions in
  liquid water revealed by two-dimensional terahertz-infrared-visible
  spectroscopy},}\ }\href {https://doi.org/10.1038/s41467-018-03303-y}
  {\bibfield  {journal} {\bibinfo  {journal} {Nat Commun}\ }\textbf {\bibinfo
  {volume} {9}},\ \bibinfo {pages} {885} (\bibinfo {year} {2018})}\BibitemShut
  {NoStop}%
\bibitem [{\citenamefont {Vietze}\ \emph {et~al.}(2021)\citenamefont {Vietze},
  \citenamefont {Backus}, \citenamefont {Bonn},\ and\ \citenamefont
  {Grechko}}]{Bonn2DTZIFvis2021}%
  \BibitemOpen
  \bibfield  {author} {\bibinfo {author} {\bibfnamefont {L.}~\bibnamefont
  {Vietze}}, \bibinfo {author} {\bibfnamefont {E.~H.~G.}\ \bibnamefont
  {Backus}}, \bibinfo {author} {\bibfnamefont {M.}~\bibnamefont {Bonn}},\ and\
  \bibinfo {author} {\bibfnamefont {M.}~\bibnamefont {Grechko}},\ }\bibfield
  {title} {\enquote {\bibinfo {title} {Distinguishing different excitation
  pathways in two-dimensional \uppercase{T}erahertz-infrared-visible
  spectroscopy},}\ }\href {https://doi.org/10.1063/5.0047918} {\bibfield
  {journal} {\bibinfo  {journal} {The Journal of Chemical Physics}\ }\textbf
  {\bibinfo {volume} {154}},\ \bibinfo {pages} {174201} (\bibinfo {year}
  {2021})},\ \Eprint {https://arxiv.org/abs/https://doi.org/10.1063/5.0047918}
  {https://doi.org/10.1063/5.0047918} \BibitemShut {NoStop}%
\bibitem [{\citenamefont {Ito}\ and\ \citenamefont {Tanimura}(2016)}]{IT16JCP}%
  \BibitemOpen
  \bibfield  {author} {\bibinfo {author} {\bibfnamefont {H.}~\bibnamefont
  {Ito}}\ and\ \bibinfo {author} {\bibfnamefont {Y.}~\bibnamefont {Tanimura}},\
  }\bibfield  {title} {\enquote {\bibinfo {title} {Simulating two-dimensional
  infrared-\uppercase{R}aman and \uppercase{R}aman spectroscopies for
  intermolecular and intramolecular modes of liquid water},}\ }\href
  {https://doi.org/10.1063/1.4941842} {\bibfield  {journal} {\bibinfo
  {journal} {The Journal of Chemical Physics}\ }\textbf {\bibinfo {volume}
  {144}},\ \bibinfo {pages} {074201} (\bibinfo {year} {2016})}\BibitemShut
  {NoStop}%
\bibitem [{\citenamefont {Takahashi}\ and\ \citenamefont
  {Tanimura}(2023)}]{TT23JCP1}%
  \BibitemOpen
  \bibfield  {author} {\bibinfo {author} {\bibfnamefont {H.}~\bibnamefont
  {Takahashi}}\ and\ \bibinfo {author} {\bibfnamefont {Y.}~\bibnamefont
  {Tanimura}},\ }\bibfield  {title} {\enquote {\bibinfo {title} {Discretized
  hierarchical equations of motion in mixed liouville–wigner space for
  two-dimensional vibrational spectroscopies of liquid water},}\ }\href
  {https://doi.org/10.1063/5.0135725} {\bibfield  {journal} {\bibinfo
  {journal} {The Journal of Chemical Physics}\ }\textbf {\bibinfo {volume}
  {158}},\ \bibinfo {pages} {044115} (\bibinfo {year} {2023})},\ \Eprint
  {https://arxiv.org/abs/https://doi.org/10.1063/5.0135725}
  {https://doi.org/10.1063/5.0135725} \BibitemShut {NoStop}%
\bibitem [{\citenamefont {Okumura}\ and\ \citenamefont
  {Tanimura}(2003)}]{OT03JPC}%
  \BibitemOpen
  \bibfield  {author} {\bibinfo {author} {\bibfnamefont {K.}~\bibnamefont
  {Okumura}}\ and\ \bibinfo {author} {\bibfnamefont {Y.}~\bibnamefont
  {Tanimura}},\ }\bibfield  {title} {\enquote {\bibinfo {title} {Energy-level
  diagrams and their contribution to fifth-order \uppercase{R}aman and
  second-order infrared responses: distinction between relaxation models by
  two-dimensional spectroscopy},}\ }\href {https://doi.org/10.1021/jp027360o}
  {\bibfield  {journal} {\bibinfo  {journal} {The Journal of Physical Chemistry
  A}\ }\textbf {\bibinfo {volume} {107}},\ \bibinfo {pages} {8092--8105}
  (\bibinfo {year} {2003})}\BibitemShut {NoStop}%
\bibitem [{\citenamefont {Sidler}\ and\ \citenamefont
  {Hamm}(2020)}]{Hamm2DTHzFeynmann}%
  \BibitemOpen
  \bibfield  {author} {\bibinfo {author} {\bibfnamefont {D.}~\bibnamefont
  {Sidler}}\ and\ \bibinfo {author} {\bibfnamefont {P.}~\bibnamefont {Hamm}},\
  }\bibfield  {title} {\enquote {\bibinfo {title} {A feynman diagram
  description of the 2d-raman-thz response of amorphous ice},}\ }\href
  {https://doi.org/10.1063/5.0018485} {\bibfield  {journal} {\bibinfo
  {journal} {The Journal of Chemical Physics}\ }\textbf {\bibinfo {volume}
  {153}},\ \bibinfo {pages} {044502} (\bibinfo {year} {2020})},\ \Eprint
  {https://arxiv.org/abs/https://doi.org/10.1063/5.0018485}
  {https://doi.org/10.1063/5.0018485} \BibitemShut {NoStop}%
\bibitem [{\citenamefont {Ito}, \citenamefont {Hasegawa},\ and\ \citenamefont
  {Tanimura}(2014)}]{IHT14JCP}%
  \BibitemOpen
  \bibfield  {author} {\bibinfo {author} {\bibfnamefont {H.}~\bibnamefont
  {Ito}}, \bibinfo {author} {\bibfnamefont {T.}~\bibnamefont {Hasegawa}},\ and\
  \bibinfo {author} {\bibfnamefont {Y.}~\bibnamefont {Tanimura}},\ }\bibfield
  {title} {\enquote {\bibinfo {title} {Calculating two-dimensional
  \uppercase{TH}z-\uppercase{R}aman-\uppercase{TH}z and
  \uppercase{R}aman-\uppercase{TH}z-\uppercase{TH}z signals for various
  molecular liquids: The samplers},}\ }\href
  {https://doi.org/10.1063/1.4895908} {\bibfield  {journal} {\bibinfo
  {journal} {The Journal of Chemical Physics}\ }\textbf {\bibinfo {volume}
  {141}},\ \bibinfo {pages} {124503} (\bibinfo {year} {2014})}\BibitemShut
  {NoStop}%
\bibitem [{\citenamefont {Ito}, \citenamefont {Jo},\ and\ \citenamefont
  {Tanimura}(2015)}]{IJT15SD}%
  \BibitemOpen
  \bibfield  {author} {\bibinfo {author} {\bibfnamefont {H.}~\bibnamefont
  {Ito}}, \bibinfo {author} {\bibfnamefont {J.-Y.}\ \bibnamefont {Jo}},\ and\
  \bibinfo {author} {\bibfnamefont {Y.}~\bibnamefont {Tanimura}},\ }\bibfield
  {title} {\enquote {\bibinfo {title} {Notes on simulating two-dimensional
  \uppercase{R}aman and terahertz-\uppercase{R}aman signals with a full
  molecular dynamics simulation approach},}\ }\href
  {https://doi.org/10.1063/1.4932597} {\bibfield  {journal} {\bibinfo
  {journal} {Structural Dynamics}\ }\textbf {\bibinfo {volume} {2}},\ \bibinfo
  {pages} {054102} (\bibinfo {year} {2015})}\BibitemShut {NoStop}%
\bibitem [{\citenamefont {Ito}, \citenamefont {Hasegawa},\ and\ \citenamefont
  {Tanimura}(2016)}]{IHT16JPCL}%
  \BibitemOpen
  \bibfield  {author} {\bibinfo {author} {\bibfnamefont {H.}~\bibnamefont
  {Ito}}, \bibinfo {author} {\bibfnamefont {T.}~\bibnamefont {Hasegawa}},\ and\
  \bibinfo {author} {\bibfnamefont {Y.}~\bibnamefont {Tanimura}},\ }\bibfield
  {title} {\enquote {\bibinfo {title} {Effects of intermolecular charge
  transfer in liquid water on \uppercase{R}aman spectra},}\ }\href
  {https://doi.org/10.1021/acs.jpclett.6b01766} {\bibfield  {journal} {\bibinfo
   {journal} {The Journal of Physical Chemistry Letters}\ }\textbf {\bibinfo
  {volume} {7}},\ \bibinfo {pages} {4147--4151} (\bibinfo {year}
  {2016})}\BibitemShut {NoStop}%
\bibitem [{\citenamefont {Ikeda}, \citenamefont {Ito},\ and\ \citenamefont
  {Tanimura}(2015)}]{IIT15JCP}%
  \BibitemOpen
  \bibfield  {author} {\bibinfo {author} {\bibfnamefont {T.}~\bibnamefont
  {Ikeda}}, \bibinfo {author} {\bibfnamefont {H.}~\bibnamefont {Ito}},\ and\
  \bibinfo {author} {\bibfnamefont {Y.}~\bibnamefont {Tanimura}},\ }\bibfield
  {title} {\enquote {\bibinfo {title} {Analysis of 2d
  \uppercase{TH}z-\uppercase{R}aman spectroscopy using a
  non-\uppercase{M}arkovian \uppercase{B}rownian oscillator model with
  nonlinear system-bath interactions},}\ }\href
  {https://doi.org/10.1063/1.4917033} {\bibfield  {journal} {\bibinfo
  {journal} {The Journal of Chemical Physics}\ }\textbf {\bibinfo {volume}
  {142}},\ \bibinfo {pages} {212421} (\bibinfo {year} {2015})}\BibitemShut
  {NoStop}%
\bibitem [{\citenamefont {Ishizaki}\ and\ \citenamefont
  {Tanimura}(2006)}]{IT06JCP}%
  \BibitemOpen
  \bibfield  {author} {\bibinfo {author} {\bibfnamefont {A.}~\bibnamefont
  {Ishizaki}}\ and\ \bibinfo {author} {\bibfnamefont {Y.}~\bibnamefont
  {Tanimura}},\ }\bibfield  {title} {\enquote {\bibinfo {title} {Modeling
  vibrational dephasing and energy relaxation of intramolecular anharmonic
  modes for multidimensional infrared spectroscopies},}\ }\href
  {https://doi.org/10.1063/1.2244558} {\bibfield  {journal} {\bibinfo
  {journal} {The Journal of Chemical Physics}\ }\textbf {\bibinfo {volume}
  {125}},\ \bibinfo {pages} {084501} (\bibinfo {year} {2006})}\BibitemShut
  {NoStop}%
\bibitem [{\citenamefont {Sakurai}\ and\ \citenamefont
  {Tanimura}(2011)}]{ST11JPCA}%
  \BibitemOpen
  \bibfield  {author} {\bibinfo {author} {\bibfnamefont {A.}~\bibnamefont
  {Sakurai}}\ and\ \bibinfo {author} {\bibfnamefont {Y.}~\bibnamefont
  {Tanimura}},\ }\bibfield  {title} {\enquote {\bibinfo {title} {Does $\hbar$
  play a role in multidimensional spectroscopy? reduced hierarchy equations of
  motion approach to molecular vibrations},}\ }\href
  {https://doi.org/10.1021/jp1095618} {\bibfield  {journal} {\bibinfo
  {journal} {The Journal of Physical Chemistry A}\ }\textbf {\bibinfo {volume}
  {115}},\ \bibinfo {pages} {4009--4022} (\bibinfo {year} {2011})}\BibitemShut
  {NoStop}%
\bibitem [{\citenamefont {Jeon}\ and\ \citenamefont
  {Cho}(2014)}]{ChoH2OMD2014}%
  \BibitemOpen
  \bibfield  {author} {\bibinfo {author} {\bibfnamefont {J.}~\bibnamefont
  {Jeon}}\ and\ \bibinfo {author} {\bibfnamefont {M.}~\bibnamefont {Cho}},\
  }\bibfield  {title} {\enquote {\bibinfo {title} {An accurate classical
  simulation of a two-dimensional vibrational spectrum: \uppercase{OD} stretch
  spectrum of a hydrated \uppercase{HOD} molecule},}\ }\href
  {https://doi.org/10.1021/jp501182d} {\bibfield  {journal} {\bibinfo
  {journal} {The Journal of Physical Chemistry B}\ }\textbf {\bibinfo {volume}
  {118}},\ \bibinfo {pages} {8148--8161} (\bibinfo {year} {2014})},\ \bibinfo
  {note} {pMID: 24601590},\ \Eprint
  {https://arxiv.org/abs/https://doi.org/10.1021/jp501182d}
  {https://doi.org/10.1021/jp501182d} \BibitemShut {NoStop}%
\bibitem [{\citenamefont {Imoto}, \citenamefont {Xantheas},\ and\ \citenamefont
  {Saito}(2013)}]{ImotXanteasSaitoJCP2013H2O}%
  \BibitemOpen
  \bibfield  {author} {\bibinfo {author} {\bibfnamefont {S.}~\bibnamefont
  {Imoto}}, \bibinfo {author} {\bibfnamefont {S.~S.}\ \bibnamefont
  {Xantheas}},\ and\ \bibinfo {author} {\bibfnamefont {S.}~\bibnamefont
  {Saito}},\ }\bibfield  {title} {\enquote {\bibinfo {title} {Ultrafast
  dynamics of liquid water: Frequency fluctuations of the \uppercase{OH}
  stretch and the \uppercase{HOH} bend},}\ }\href
  {https://doi.org/10.1063/1.4813071} {\bibfield  {journal} {\bibinfo
  {journal} {The Journal of Chemical Physics}\ }\textbf {\bibinfo {volume}
  {139}},\ \bibinfo {pages} {044503} (\bibinfo {year} {2013})},\ \Eprint
  {https://arxiv.org/abs/https://doi.org/10.1063/1.4813071}
  {https://doi.org/10.1063/1.4813071} \BibitemShut {NoStop}%
\bibitem [{\citenamefont {Piryatinski}, \citenamefont {Lawrence},\ and\
  \citenamefont {Skinner}(2003)}]{SkinnerStochs2003}%
  \BibitemOpen
  \bibfield  {author} {\bibinfo {author} {\bibfnamefont {A.}~\bibnamefont
  {Piryatinski}}, \bibinfo {author} {\bibfnamefont {C.~P.}\ \bibnamefont
  {Lawrence}},\ and\ \bibinfo {author} {\bibfnamefont {J.~L.}\ \bibnamefont
  {Skinner}},\ }\bibfield  {title} {\enquote {\bibinfo {title} {Vibrational
  spectroscopy of \uppercase{HOD} in liquid \uppercase{D}$_2$\uppercase{O}. v.
  infrared three-pulse photon echoes},}\ }\href
  {https://doi.org/10.1063/1.1569474} {\bibfield  {journal} {\bibinfo
  {journal} {The Journal of Chemical Physics}\ }\textbf {\bibinfo {volume}
  {118}},\ \bibinfo {pages} {9672--9679} (\bibinfo {year} {2003})},\ \Eprint
  {https://arxiv.org/abs/https://doi.org/10.1063/1.1569474}
  {https://doi.org/10.1063/1.1569474} \BibitemShut {NoStop}%
\bibitem [{\citenamefont {Paarmann}\ \emph {et~al.}(2009)\citenamefont
  {Paarmann}, \citenamefont {Hayashi}, \citenamefont {Mukamel},\ and\
  \citenamefont {Miller}}]{Mukamel2009}%
  \BibitemOpen
  \bibfield  {author} {\bibinfo {author} {\bibfnamefont {A.}~\bibnamefont
  {Paarmann}}, \bibinfo {author} {\bibfnamefont {T.}~\bibnamefont {Hayashi}},
  \bibinfo {author} {\bibfnamefont {S.}~\bibnamefont {Mukamel}},\ and\ \bibinfo
  {author} {\bibfnamefont {R.~J.~D.}\ \bibnamefont {Miller}},\ }\bibfield
  {title} {\enquote {\bibinfo {title} {Nonlinear response of vibrational
  excitons: Simulating the two-dimensional infrared spectrum of liquid
  water},}\ }\href {https://doi.org/10.1063/1.3139003} {\bibfield  {journal}
  {\bibinfo  {journal} {The Journal of Chemical Physics}\ }\textbf {\bibinfo
  {volume} {130}},\ \bibinfo {pages} {204110} (\bibinfo {year} {2009})},\
  \Eprint {https://arxiv.org/abs/https://doi.org/10.1063/1.3139003}
  {https://doi.org/10.1063/1.3139003} \BibitemShut {NoStop}%
\bibitem [{\citenamefont {Jansen}\ \emph {et~al.}(2010)\citenamefont {Jansen},
  \citenamefont {Auer}, \citenamefont {Yang},\ and\ \citenamefont
  {Skinner}}]{JansenSkiner2010}%
  \BibitemOpen
  \bibfield  {author} {\bibinfo {author} {\bibfnamefont {T.~l.~C.}\
  \bibnamefont {Jansen}}, \bibinfo {author} {\bibfnamefont {B.~M.}\
  \bibnamefont {Auer}}, \bibinfo {author} {\bibfnamefont {M.}~\bibnamefont
  {Yang}},\ and\ \bibinfo {author} {\bibfnamefont {J.~L.}\ \bibnamefont
  {Skinner}},\ }\bibfield  {title} {\enquote {\bibinfo {title} {Two-dimensional
  infrared spectroscopy and ultrafast anisotropy decay of water},}\ }\href
  {https://doi.org/10.1063/1.3454733} {\bibfield  {journal} {\bibinfo
  {journal} {The Journal of Chemical Physics}\ }\textbf {\bibinfo {volume}
  {132}},\ \bibinfo {pages} {224503} (\bibinfo {year} {2010})}\BibitemShut
  {NoStop}%
\bibitem [{\citenamefont {Gottwald}, \citenamefont {Ivanov},\ and\
  \citenamefont {Kühn}(2015)}]{Kuehn2015JPCL}%
  \BibitemOpen
  \bibfield  {author} {\bibinfo {author} {\bibfnamefont {F.}~\bibnamefont
  {Gottwald}}, \bibinfo {author} {\bibfnamefont {S.~D.}\ \bibnamefont
  {Ivanov}},\ and\ \bibinfo {author} {\bibfnamefont {O.}~\bibnamefont
  {Kühn}},\ }\bibfield  {title} {\enquote {\bibinfo {title} {Applicability of
  the \uppercase{C}aldeira–\uppercase{L}eggett model to vibrational
  spectroscopy in solution},}\ }\href
  {https://doi.org/10.1021/acs.jpclett.5b00718} {\bibfield  {journal} {\bibinfo
   {journal} {The Journal of Physical Chemistry Letters}\ }\textbf {\bibinfo
  {volume} {6}},\ \bibinfo {pages} {2722--2727} (\bibinfo {year}
  {2015})}\BibitemShut {NoStop}%
\bibitem [{\citenamefont {Gottwald}\ \emph {et~al.}(2015)\citenamefont
  {Gottwald}, \citenamefont {Karsten}, \citenamefont {Ivanov},\ and\
  \citenamefont {Kühn}}]{Kuehn2015JCP}%
  \BibitemOpen
  \bibfield  {author} {\bibinfo {author} {\bibfnamefont {F.}~\bibnamefont
  {Gottwald}}, \bibinfo {author} {\bibfnamefont {S.}~\bibnamefont {Karsten}},
  \bibinfo {author} {\bibfnamefont {S.~D.}\ \bibnamefont {Ivanov}},\ and\
  \bibinfo {author} {\bibfnamefont {O.}~\bibnamefont {Kühn}},\ }\bibfield
  {title} {\enquote {\bibinfo {title} {Parametrizing linear generalized
  \uppercase{L}angevin dynamics from explicit molecular dynamics
  simulations},}\ }\href {https://doi.org/10.1063/1.4922941} {\bibfield
  {journal} {\bibinfo  {journal} {The Journal of Chemical Physics}\ }\textbf
  {\bibinfo {volume} {142}},\ \bibinfo {pages} {244110} (\bibinfo {year}
  {2015})}\BibitemShut {NoStop}%
\bibitem [{\citenamefont {Gottwald}, \citenamefont {Ivanov},\ and\
  \citenamefont {Kühn}(2016)}]{Kuehn2016JCP}%
  \BibitemOpen
  \bibfield  {author} {\bibinfo {author} {\bibfnamefont {F.}~\bibnamefont
  {Gottwald}}, \bibinfo {author} {\bibfnamefont {S.~D.}\ \bibnamefont
  {Ivanov}},\ and\ \bibinfo {author} {\bibfnamefont {O.}~\bibnamefont
  {Kühn}},\ }\bibfield  {title} {\enquote {\bibinfo {title} {Vibrational
  spectroscopy via the \uppercase{C}aldeira–\uppercase{L}eggett model with
  anharmonic system potentials},}\ }\href {https://doi.org/10.1063/1.4946872}
  {\bibfield  {journal} {\bibinfo  {journal} {The Journal of Chemical Physics}\
  }\textbf {\bibinfo {volume} {144}},\ \bibinfo {pages} {164102} (\bibinfo
  {year} {2016})}\BibitemShut {NoStop}%
\bibitem [{\citenamefont {Yagasaki}, \citenamefont {Ono},\ and\ \citenamefont
  {Saito}(2009)}]{Yagasakirelax2009}%
  \BibitemOpen
  \bibfield  {author} {\bibinfo {author} {\bibfnamefont {T.}~\bibnamefont
  {Yagasaki}}, \bibinfo {author} {\bibfnamefont {J.}~\bibnamefont {Ono}},\ and\
  \bibinfo {author} {\bibfnamefont {S.}~\bibnamefont {Saito}},\ }\bibfield
  {title} {\enquote {\bibinfo {title} {Ultrafast energy relaxation and
  anisotropy decay of the librational motion in liquid water: A molecular
  dynamics study},}\ }\href {https://doi.org/10.1063/1.3254518} {\bibfield
  {journal} {\bibinfo  {journal} {The Journal of Chemical Physics}\ }\textbf
  {\bibinfo {volume} {131}},\ \bibinfo {pages} {164511} (\bibinfo {year}
  {2009})},\ \Eprint {https://arxiv.org/abs/https://doi.org/10.1063/1.3254518}
  {https://doi.org/10.1063/1.3254518} \BibitemShut {NoStop}%
\bibitem [{\citenamefont {Yagasaki}\ and\ \citenamefont
  {Saito}(2011)}]{YagasakiSaitoJCP2011Relax}%
  \BibitemOpen
  \bibfield  {author} {\bibinfo {author} {\bibfnamefont {T.}~\bibnamefont
  {Yagasaki}}\ and\ \bibinfo {author} {\bibfnamefont {S.}~\bibnamefont
  {Saito}},\ }\bibfield  {title} {\enquote {\bibinfo {title} {A novel method
  for analyzing energy relaxation in condensed phases using nonequilibrium
  molecular dynamics simulations: Application to the energy relaxation of
  intermolecular motions in liquid water},}\ }\href
  {https://doi.org/10.1063/1.3587105} {\bibfield  {journal} {\bibinfo
  {journal} {The Journal of Chemical Physics}\ }\textbf {\bibinfo {volume}
  {134}},\ \bibinfo {pages} {184503} (\bibinfo {year} {2011})},\ \Eprint
  {https://arxiv.org/abs/https://doi.org/10.1063/1.3587105}
  {https://doi.org/10.1063/1.3587105} \BibitemShut {NoStop}%
\bibitem [{\citenamefont {Okumura}\ and\ \citenamefont
  {Tanimura}(1997{\natexlab{a}})}]{OT97PRE}%
  \BibitemOpen
  \bibfield  {author} {\bibinfo {author} {\bibfnamefont {K.}~\bibnamefont
  {Okumura}}\ and\ \bibinfo {author} {\bibfnamefont {Y.}~\bibnamefont
  {Tanimura}},\ }\bibfield  {title} {\enquote {\bibinfo {title} {Two-time
  correlation functions of a harmonic system nonbilinearly coupled to a heat
  bath: Spontaneous \uppercase{R}aman spectroscopy},}\ }\href
  {https://doi.org/10.1103/PhysRevE.56.2747} {\bibfield  {journal} {\bibinfo
  {journal} {Phys. Rev. E}\ }\textbf {\bibinfo {volume} {56}},\ \bibinfo
  {pages} {2747--2750} (\bibinfo {year} {1997}{\natexlab{a}})}\BibitemShut
  {NoStop}%
\bibitem [{\citenamefont {Tanimura}\ and\ \citenamefont
  {Steffen}(2000)}]{TS20JPSJ}%
  \BibitemOpen
  \bibfield  {author} {\bibinfo {author} {\bibfnamefont {Y.}~\bibnamefont
  {Tanimura}}\ and\ \bibinfo {author} {\bibfnamefont {T.}~\bibnamefont
  {Steffen}},\ }\bibfield  {title} {\enquote {\bibinfo {title} {Two-dimensional
  spectroscopy for harmonic vibrational modes with nonlinear system-bath
  interactions.ii. \uppercase{G}aussian-\uppercase{M}arkovian case},}\ }\href
  {https://doi.org/10.1143/JPSJ.69.4095} {\bibfield  {journal} {\bibinfo
  {journal} {Journal of the Physical Society of Japan}\ }\textbf {\bibinfo
  {volume} {69}},\ \bibinfo {pages} {4095--4106} (\bibinfo {year}
  {2000})}\BibitemShut {NoStop}%
\bibitem [{\citenamefont {Kato}\ and\ \citenamefont
  {Tanimura}(2004)}]{KT04JCP}%
  \BibitemOpen
  \bibfield  {author} {\bibinfo {author} {\bibfnamefont {T.}~\bibnamefont
  {Kato}}\ and\ \bibinfo {author} {\bibfnamefont {Y.}~\bibnamefont
  {Tanimura}},\ }\bibfield  {title} {\enquote {\bibinfo {title}
  {Two-dimensional \uppercase{R}aman and infrared vibrational spectroscopy for
  a harmonic oscillator system nonlinearly coupled with a colored noise
  bath},}\ }\href {https://doi.org/10.1063/1.1629272} {\bibfield  {journal}
  {\bibinfo  {journal} {The Journal of Chemical Physics}\ }\textbf {\bibinfo
  {volume} {120}},\ \bibinfo {pages} {260--271} (\bibinfo {year}
  {2004})}\BibitemShut {NoStop}%
\bibitem [{\citenamefont {Ishizaki}\ and\ \citenamefont
  {Tanimura}(2007)}]{IT07JPCA}%
  \BibitemOpen
  \bibfield  {author} {\bibinfo {author} {\bibfnamefont {A.}~\bibnamefont
  {Ishizaki}}\ and\ \bibinfo {author} {\bibfnamefont {Y.}~\bibnamefont
  {Tanimura}},\ }\bibfield  {title} {\enquote {\bibinfo {title} {Dynamics of a
  multimode system coupled to multiple heat baths probed by two-dimensional
  infrared spectroscopy},}\ }\href {https://doi.org/10.1021/jp072880a}
  {\bibfield  {journal} {\bibinfo  {journal} {The Journal of Physical Chemistry
  A}\ }\textbf {\bibinfo {volume} {111}},\ \bibinfo {pages} {9269--9276}
  (\bibinfo {year} {2007})}\BibitemShut {NoStop}%
\bibitem [{\citenamefont {Tanimura}\ and\ \citenamefont
  {Kubo}(1989)}]{TK89JPSJ1}%
  \BibitemOpen
  \bibfield  {author} {\bibinfo {author} {\bibfnamefont {Y.}~\bibnamefont
  {Tanimura}}\ and\ \bibinfo {author} {\bibfnamefont {R.}~\bibnamefont
  {Kubo}},\ }\bibfield  {title} {\enquote {\bibinfo {title} {Time evolution of
  a quantum system in contact with a nearly
  \uppercase{G}aussian-\uppercase{M}arkoffian noise bath},}\ }\href
  {https://doi.org/10.1143/JPSJ.58.101} {\bibfield  {journal} {\bibinfo
  {journal} {Journal of the Physical Society of Japan}\ }\textbf {\bibinfo
  {volume} {58}},\ \bibinfo {pages} {101--114} (\bibinfo {year}
  {1989})}\BibitemShut {NoStop}%
\bibitem [{\citenamefont {Tanimura}(2006)}]{T06JPSJ}%
  \BibitemOpen
  \bibfield  {author} {\bibinfo {author} {\bibfnamefont {Y.}~\bibnamefont
  {Tanimura}},\ }\bibfield  {title} {\enquote {\bibinfo {title} {Stochastic
  \uppercase{L}iouville, \uppercase{L}angevin,
  \uppercase{F}okker-\uppercase{P}lanck, and master equation qpproaches to
  quantum dissipative systems},}\ }\href
  {https://doi.org/10.1143/JPSJ.75.082001} {\bibfield  {journal} {\bibinfo
  {journal} {Journal of the Physical Society of Japan}\ }\textbf {\bibinfo
  {volume} {75}},\ \bibinfo {pages} {082001} (\bibinfo {year}
  {2006})}\BibitemShut {NoStop}%
\bibitem [{\citenamefont {Tanimura}(2020)}]{T20JCP}%
  \BibitemOpen
  \bibfield  {author} {\bibinfo {author} {\bibfnamefont {Y.}~\bibnamefont
  {Tanimura}},\ }\bibfield  {title} {\enquote {\bibinfo {title} {Numerically
  "exact" approach to open quantum dynamics: The hierarchical equations of
  motion (\uppercase{HEOM})},}\ }\href {https://doi.org/10.1063/5.0011599}
  {\bibfield  {journal} {\bibinfo  {journal} {The Journal of Chemical Physics}\
  }\textbf {\bibinfo {volume} {153}},\ \bibinfo {pages} {020901} (\bibinfo
  {year} {2020})}\BibitemShut {NoStop}%
\bibitem [{\citenamefont {Hasegawa}\ and\ \citenamefont
  {Tanimura}(2011)}]{HT11JPCB}%
  \BibitemOpen
  \bibfield  {author} {\bibinfo {author} {\bibfnamefont {T.}~\bibnamefont
  {Hasegawa}}\ and\ \bibinfo {author} {\bibfnamefont {Y.}~\bibnamefont
  {Tanimura}},\ }\bibfield  {title} {\enquote {\bibinfo {title} {A polarizable
  water model for intramolecular and intermolecular vibrational
  spectroscopies},}\ }\href {https://doi.org/10.1021/jp111308f} {\bibfield
  {journal} {\bibinfo  {journal} {The Journal of Physical Chemistry B}\
  }\textbf {\bibinfo {volume} {115}},\ \bibinfo {pages} {5545--5553} (\bibinfo
  {year} {2011})}\BibitemShut {NoStop}%
\bibitem [{\citenamefont {Liu}\ and\ \citenamefont
  {Liu}(2018)}]{JianLiu2018H2OMP}%
  \BibitemOpen
  \bibfield  {author} {\bibinfo {author} {\bibfnamefont {X.}~\bibnamefont
  {Liu}}\ and\ \bibinfo {author} {\bibfnamefont {J.}~\bibnamefont {Liu}},\
  }\bibfield  {title} {\enquote {\bibinfo {title} {Critical role of quantum
  dynamical effects in the \uppercase{R}aman spectroscopy of liquid water},}\
  }\href {https://doi.org/10.1080/00268976.2018.1434907} {\bibfield  {journal}
  {\bibinfo  {journal} {Molecular Physics}\ }\textbf {\bibinfo {volume}
  {116}},\ \bibinfo {pages} {755--779} (\bibinfo {year} {2018})},\ \Eprint
  {https://arxiv.org/abs/https://doi.org/10.1080/00268976.2018.1434907}
  {https://doi.org/10.1080/00268976.2018.1434907} \BibitemShut {NoStop}%
\bibitem [{\citenamefont {Liu}, \citenamefont {Wang},\ and\ \citenamefont
  {Bowman}(2015)}]{BowmanQM_QD2015}%
  \BibitemOpen
  \bibfield  {author} {\bibinfo {author} {\bibfnamefont {H.}~\bibnamefont
  {Liu}}, \bibinfo {author} {\bibfnamefont {Y.}~\bibnamefont {Wang}},\ and\
  \bibinfo {author} {\bibfnamefont {J.~M.}\ \bibnamefont {Bowman}},\ }\bibfield
   {title} {\enquote {\bibinfo {title} {Quantum calculations of the ir spectrum
  of liquid water using ab initio and model potential and dipole moment
  surfaces and comparison with experiment},}\ }\href
  {https://doi.org/10.1063/1.4921045} {\bibfield  {journal} {\bibinfo
  {journal} {The Journal of Chemical Physics}\ }\textbf {\bibinfo {volume}
  {142}},\ \bibinfo {pages} {194502} (\bibinfo {year} {2015})},\ \Eprint
  {https://arxiv.org/abs/https://doi.org/10.1063/1.4921045}
  {https://doi.org/10.1063/1.4921045} \BibitemShut {NoStop}%
\bibitem [{\citenamefont {Hunter}, \citenamefont {Shakib},\ and\ \citenamefont
  {Paesani}(2018)}]{Paesan2018H2OCMD}%
  \BibitemOpen
  \bibfield  {author} {\bibinfo {author} {\bibfnamefont {K.~M.}\ \bibnamefont
  {Hunter}}, \bibinfo {author} {\bibfnamefont {F.~A.}\ \bibnamefont {Shakib}},\
  and\ \bibinfo {author} {\bibfnamefont {F.}~\bibnamefont {Paesani}},\
  }\bibfield  {title} {\enquote {\bibinfo {title} {Disentangling coupling
  effects in the infrared spectra of liquid water},}\ }\href
  {https://doi.org/10.1021/acs.jpcb.8b09910} {\bibfield  {journal} {\bibinfo
  {journal} {The Journal of Physical Chemistry B}\ }\textbf {\bibinfo {volume}
  {122}},\ \bibinfo {pages} {10754--10761} (\bibinfo {year} {2018})},\ \bibinfo
  {note} {pMID: 30403350},\ \Eprint
  {https://arxiv.org/abs/https://doi.org/10.1021/acs.jpcb.8b09910}
  {https://doi.org/10.1021/acs.jpcb.8b09910} \BibitemShut {NoStop}%
\bibitem [{\citenamefont {Trenins}, \citenamefont {Willatt},\ and\
  \citenamefont {Althorpe}(2019)}]{Althorpe2019CMD}%
  \BibitemOpen
  \bibfield  {author} {\bibinfo {author} {\bibfnamefont {G.}~\bibnamefont
  {Trenins}}, \bibinfo {author} {\bibfnamefont {M.~J.}\ \bibnamefont
  {Willatt}},\ and\ \bibinfo {author} {\bibfnamefont {S.~C.}\ \bibnamefont
  {Althorpe}},\ }\bibfield  {title} {\enquote {\bibinfo {title} {Path-integral
  dynamics of water using curvilinear centroids},}\ }\href
  {https://doi.org/10.1063/1.5100587} {\bibfield  {journal} {\bibinfo
  {journal} {The Journal of Chemical Physics}\ }\textbf {\bibinfo {volume}
  {151}},\ \bibinfo {pages} {054109} (\bibinfo {year} {2019})},\ \Eprint
  {https://arxiv.org/abs/https://doi.org/10.1063/1.5100587}
  {https://doi.org/10.1063/1.5100587} \BibitemShut {NoStop}%
\bibitem [{\citenamefont {Ueno}\ and\ \citenamefont
  {Tanimura}(2020)}]{UT20JCTC}%
  \BibitemOpen
  \bibfield  {author} {\bibinfo {author} {\bibfnamefont {S.}~\bibnamefont
  {Ueno}}\ and\ \bibinfo {author} {\bibfnamefont {Y.}~\bibnamefont
  {Tanimura}},\ }\bibfield  {title} {\enquote {\bibinfo {title} {Modeling
  intermolecular and intramolecular modes of liquid water using multiple heat
  baths: Machine learning approach},}\ }\href
  {https://doi.org/10.1021/acs.jctc.9b01288} {\bibfield  {journal} {\bibinfo
  {journal} {Journal of Chemical Theory and Computation}\ }\textbf {\bibinfo
  {volume} {16}},\ \bibinfo {pages} {2099--2108} (\bibinfo {year}
  {2020})}\BibitemShut {NoStop}%
\bibitem [{\citenamefont {Okumura}\ and\ \citenamefont
  {Tanimura}(1997{\natexlab{b}})}]{OT97JCP1}%
  \BibitemOpen
  \bibfield  {author} {\bibinfo {author} {\bibfnamefont {K.}~\bibnamefont
  {Okumura}}\ and\ \bibinfo {author} {\bibfnamefont {Y.}~\bibnamefont
  {Tanimura}},\ }\bibfield  {title} {\enquote {\bibinfo {title} {The
  (2n+1)th-order off-resonant spectroscopy from the (n+1)th-order
  anharmonicities of molecular vibrational modes in the condensed phase},}\
  }\href {https://doi.org/10.1063/1.473284} {\bibfield  {journal} {\bibinfo
  {journal} {The Journal of Chemical Physics}\ }\textbf {\bibinfo {volume}
  {106}},\ \bibinfo {pages} {1687--1698} (\bibinfo {year}
  {1997}{\natexlab{b}})}\BibitemShut {NoStop}%
\bibitem [{\citenamefont {Okumura}\ and\ \citenamefont
  {Tanimura}(1997{\natexlab{c}})}]{OT97JCP2}%
  \BibitemOpen
  \bibfield  {author} {\bibinfo {author} {\bibfnamefont {K.}~\bibnamefont
  {Okumura}}\ and\ \bibinfo {author} {\bibfnamefont {Y.}~\bibnamefont
  {Tanimura}},\ }\bibfield  {title} {\enquote {\bibinfo {title} {Femtosecond
  two-dimensional spectroscopy from anharmonic vibrational modes of molecules
  in the condensed phase},}\ }\href {https://doi.org/10.1063/1.474604}
  {\bibfield  {journal} {\bibinfo  {journal} {The Journal of Chemical Physics}\
  }\textbf {\bibinfo {volume} {107}},\ \bibinfo {pages} {2267--2283} (\bibinfo
  {year} {1997}{\natexlab{c}})}\BibitemShut {NoStop}%
\bibitem [{\citenamefont {Okumura}\ and\ \citenamefont
  {Tanimura}(1997{\natexlab{d}})}]{OT97CPL2}%
  \BibitemOpen
  \bibfield  {author} {\bibinfo {author} {\bibfnamefont {K.}~\bibnamefont
  {Okumura}}\ and\ \bibinfo {author} {\bibfnamefont {Y.}~\bibnamefont
  {Tanimura}},\ }\bibfield  {title} {\enquote {\bibinfo {title} {Sensitivity of
  two-dimensional fifth-order \uppercase{R}aman response to the mechanism of
  vibrational mode-mode coupling in liquid molecules},}\ }\href
  {https://doi.org/https://doi.org/10.1016/S0009-2614(97)00942-1} {\bibfield
  {journal} {\bibinfo  {journal} {Chemical Physics Letters}\ }\textbf {\bibinfo
  {volume} {278}},\ \bibinfo {pages} {175--183} (\bibinfo {year}
  {1997}{\natexlab{d}})}\BibitemShut {NoStop}%
\bibitem [{\citenamefont {Okumura}, \citenamefont {Jonas},\ and\ \citenamefont
  {Tanimura}(2001)}]{OJT01CP}%
  \BibitemOpen
  \bibfield  {author} {\bibinfo {author} {\bibfnamefont {K.}~\bibnamefont
  {Okumura}}, \bibinfo {author} {\bibfnamefont {D.~M.}\ \bibnamefont {Jonas}},\
  and\ \bibinfo {author} {\bibfnamefont {Y.}~\bibnamefont {Tanimura}},\
  }\bibfield  {title} {\enquote {\bibinfo {title} {Two-dimensional spectroscopy
  and harmonically coupled anharmonic oscillators},}\ }\href
  {https://doi.org/https://doi.org/10.1016/S0301-0104(01)00252-X} {\bibfield
  {journal} {\bibinfo  {journal} {Chemical Physics}\ }\textbf {\bibinfo
  {volume} {266}},\ \bibinfo {pages} {237--250} (\bibinfo {year}
  {2001})}\BibitemShut {NoStop}%
\bibitem [{\citenamefont {Schmidt}, \citenamefont {Corcelli},\ and\
  \citenamefont {Skinner}(2005)}]{Skinner_nonCondon2005}%
  \BibitemOpen
  \bibfield  {author} {\bibinfo {author} {\bibfnamefont {J.~R.}\ \bibnamefont
  {Schmidt}}, \bibinfo {author} {\bibfnamefont {S.~A.}\ \bibnamefont
  {Corcelli}},\ and\ \bibinfo {author} {\bibfnamefont {J.~L.}\ \bibnamefont
  {Skinner}},\ }\bibfield  {title} {\enquote {\bibinfo {title} {Pronounced
  non-condon effects in the ultrafast infrared spectroscopy of water},}\ }\href
  {https://doi.org/10.1063/1.1961472} {\bibfield  {journal} {\bibinfo
  {journal} {The Journal of Chemical Physics}\ }\textbf {\bibinfo {volume}
  {123}},\ \bibinfo {pages} {044513} (\bibinfo {year} {2005})},\ \Eprint
  {https://arxiv.org/abs/https://doi.org/10.1063/1.1961472}
  {https://doi.org/10.1063/1.1961472} \BibitemShut {NoStop}%
\bibitem [{\citenamefont {Palese}\ \emph {et~al.}(1994)\citenamefont {Palese},
  \citenamefont {Buontempo}, \citenamefont {Schilling}, \citenamefont
  {Lotshaw}, \citenamefont {Tanimura}, \citenamefont {Mukamel},\ and\
  \citenamefont {Miller}}]{PBSLTM94JPC}%
  \BibitemOpen
  \bibfield  {author} {\bibinfo {author} {\bibfnamefont {S.}~\bibnamefont
  {Palese}}, \bibinfo {author} {\bibfnamefont {J.~T.}\ \bibnamefont
  {Buontempo}}, \bibinfo {author} {\bibfnamefont {L.}~\bibnamefont
  {Schilling}}, \bibinfo {author} {\bibfnamefont {W.~T.}\ \bibnamefont
  {Lotshaw}}, \bibinfo {author} {\bibfnamefont {Y.}~\bibnamefont {Tanimura}},
  \bibinfo {author} {\bibfnamefont {S.}~\bibnamefont {Mukamel}},\ and\ \bibinfo
  {author} {\bibfnamefont {R.~J.~D.}\ \bibnamefont {Miller}},\ }\bibfield
  {title} {\enquote {\bibinfo {title} {Femtosecond two-dimensional
  \uppercase{R}aman spectroscopy of liquid water},}\ }\href
  {https://doi.org/10.1021/j100099a003} {\bibfield  {journal} {\bibinfo
  {journal} {The Journal of Physical Chemistry}\ }\textbf {\bibinfo {volume}
  {98}},\ \bibinfo {pages} {12466--12470} (\bibinfo {year} {1994})}\BibitemShut
  {NoStop}%
\bibitem [{\citenamefont {Hybl}, \citenamefont {Albrecht~Ferro},\ and\
  \citenamefont {Jonas}(2001)}]{2DCrrJonas2001}%
  \BibitemOpen
  \bibfield  {author} {\bibinfo {author} {\bibfnamefont {J.~D.}\ \bibnamefont
  {Hybl}}, \bibinfo {author} {\bibfnamefont {A.}~\bibnamefont
  {Albrecht~Ferro}},\ and\ \bibinfo {author} {\bibfnamefont {D.~M.}\
  \bibnamefont {Jonas}},\ }\bibfield  {title} {\enquote {\bibinfo {title}
  {Two-dimensional fourier transform electronic spectroscopy},}\ }\href
  {https://doi.org/10.1063/1.1398579} {\bibfield  {journal} {\bibinfo
  {journal} {The Journal of Chemical Physics}\ }\textbf {\bibinfo {volume}
  {115}},\ \bibinfo {pages} {6606--6622} (\bibinfo {year} {2001})},\ \Eprint
  {https://arxiv.org/abs/https://doi.org/10.1063/1.1398579}
  {https://doi.org/10.1063/1.1398579} \BibitemShut {NoStop}%
\bibitem [{\citenamefont {Ge}, \citenamefont {Zanni},\ and\ \citenamefont
  {Hochstrasser}(2002)}]{2DCrrGe2002}%
  \BibitemOpen
  \bibfield  {author} {\bibinfo {author} {\bibfnamefont {N.-H.}\ \bibnamefont
  {Ge}}, \bibinfo {author} {\bibfnamefont {M.~T.}\ \bibnamefont {Zanni}},\ and\
  \bibinfo {author} {\bibfnamefont {R.~M.}\ \bibnamefont {Hochstrasser}},\
  }\bibfield  {title} {\enquote {\bibinfo {title} {Effects of vibrational
  frequency correlations on two-dimensional infrared spectra†},}\ }\href@noop
  {} {\bibfield  {journal} {\bibinfo  {journal} {Journal of Physical Chemistry
  A}\ }\textbf {\bibinfo {volume} {106}},\ \bibinfo {pages} {962--972}
  (\bibinfo {year} {2002})}\BibitemShut {NoStop}%
\bibitem [{\citenamefont {Khalil}, \citenamefont {Demird\"oven},\ and\
  \citenamefont {Tokmakoff}(2003)}]{2DCrrTokmakoff2003}%
  \BibitemOpen
  \bibfield  {author} {\bibinfo {author} {\bibfnamefont {M.}~\bibnamefont
  {Khalil}}, \bibinfo {author} {\bibfnamefont {N.}~\bibnamefont
  {Demird\"oven}},\ and\ \bibinfo {author} {\bibfnamefont {A.}~\bibnamefont
  {Tokmakoff}},\ }\bibfield  {title} {\enquote {\bibinfo {title} {Obtaining
  absorptive line shapes in two-dimensional infrared vibrational correlation
  spectra},}\ }\href {https://doi.org/10.1103/PhysRevLett.90.047401} {\bibfield
   {journal} {\bibinfo  {journal} {Phys. Rev. Lett.}\ }\textbf {\bibinfo
  {volume} {90}},\ \bibinfo {pages} {047401} (\bibinfo {year}
  {2003})}\BibitemShut {NoStop}%
\bibitem [{\citenamefont {Hasegawa}\ and\ \citenamefont
  {Tanimura}(2008)}]{HT08JCP}%
  \BibitemOpen
  \bibfield  {author} {\bibinfo {author} {\bibfnamefont {T.}~\bibnamefont
  {Hasegawa}}\ and\ \bibinfo {author} {\bibfnamefont {Y.}~\bibnamefont
  {Tanimura}},\ }\bibfield  {title} {\enquote {\bibinfo {title} {Nonequilibrium
  molecular dynamics simulations with a backward-forward trajectories sampling
  for multidimensional infrared spectroscopy of molecular vibrational modes},}\
  }\href {https://doi.org/10.1063/1.2828189} {\bibfield  {journal} {\bibinfo
  {journal} {The Journal of Chemical Physics}\ }\textbf {\bibinfo {volume}
  {128}},\ \bibinfo {pages} {064511} (\bibinfo {year} {2008})}\BibitemShut
  {NoStop}%
\bibitem [{\citenamefont {Yagasaki}\ and\ \citenamefont
  {Saito}(2008)}]{YagasakiSaitoJCP20082DIR}%
  \BibitemOpen
  \bibfield  {author} {\bibinfo {author} {\bibfnamefont {T.}~\bibnamefont
  {Yagasaki}}\ and\ \bibinfo {author} {\bibfnamefont {S.}~\bibnamefont
  {Saito}},\ }\bibfield  {title} {\enquote {\bibinfo {title} {Ultrafast
  intermolecular dynamics of liquid water: A theoretical study on
  two-dimensional infrared spectroscopy},}\ }\href
  {https://doi.org/10.1063/1.2903470} {\bibfield  {journal} {\bibinfo
  {journal} {The Journal of Chemical Physics}\ }\textbf {\bibinfo {volume}
  {128}},\ \bibinfo {pages} {154521} (\bibinfo {year} {2008})},\ \Eprint
  {https://arxiv.org/abs/https://doi.org/10.1063/1.2903470}
  {https://doi.org/10.1063/1.2903470} \BibitemShut {NoStop}%
\bibitem [{\citenamefont {Kato}\ and\ \citenamefont
  {Tanimura}(2001)}]{KT01CPL}%
  \BibitemOpen
  \bibfield  {author} {\bibinfo {author} {\bibfnamefont {T.}~\bibnamefont
  {Kato}}\ and\ \bibinfo {author} {\bibfnamefont {Y.}~\bibnamefont
  {Tanimura}},\ }\bibfield  {title} {\enquote {\bibinfo {title}
  {Multi-dimensional vibrational spectroscopy measured from different
  phase-matching conditions},}\ }\href
  {https://doi.org/https://doi.org/10.1016/S0009-2614(01)00466-3} {\bibfield
  {journal} {\bibinfo  {journal} {Chemical Physics Letters}\ }\textbf {\bibinfo
  {volume} {341}},\ \bibinfo {pages} {329--337} (\bibinfo {year}
  {2001})}\BibitemShut {NoStop}%
\bibitem [{\citenamefont {Lazonder}, \citenamefont {Pshenichnikov},\ and\
  \citenamefont {Wiersma}(2006)}]{Wiersma2006}%
  \BibitemOpen
  \bibfield  {author} {\bibinfo {author} {\bibfnamefont {K.}~\bibnamefont
  {Lazonder}}, \bibinfo {author} {\bibfnamefont {M.~S.}\ \bibnamefont
  {Pshenichnikov}},\ and\ \bibinfo {author} {\bibfnamefont {D.~A.}\
  \bibnamefont {Wiersma}},\ }\bibfield  {title} {\enquote {\bibinfo {title}
  {Easy interpretation of optical two-dimensional correlation spectra},}\
  }\href {https://doi.org/10.1364/OL.31.003354} {\bibfield  {journal} {\bibinfo
   {journal} {Opt. Lett.}\ }\textbf {\bibinfo {volume} {31}},\ \bibinfo {pages}
  {3354--3356} (\bibinfo {year} {2006})}\BibitemShut {NoStop}%
\bibitem [{\citenamefont {Nienhuys}\ \emph {et~al.}(1999)\citenamefont
  {Nienhuys}, \citenamefont {Woutersen}, \citenamefont {van Santen},\ and\
  \citenamefont {Bakker}}]{Bakker1999}%
  \BibitemOpen
  \bibfield  {author} {\bibinfo {author} {\bibfnamefont {H.-K.}\ \bibnamefont
  {Nienhuys}}, \bibinfo {author} {\bibfnamefont {S.}~\bibnamefont {Woutersen}},
  \bibinfo {author} {\bibfnamefont {R.~A.}\ \bibnamefont {van Santen}},\ and\
  \bibinfo {author} {\bibfnamefont {H.~J.}\ \bibnamefont {Bakker}},\ }\bibfield
   {title} {\enquote {\bibinfo {title} {Mechanism for vibrational relaxation in
  water investigated by femtosecond infrared spectroscopy},}\ }\href
  {https://doi.org/10.1063/1.479408} {\bibfield  {journal} {\bibinfo  {journal}
  {The Journal of Chemical Physics}\ }\textbf {\bibinfo {volume} {111}},\
  \bibinfo {pages} {1494--1500} (\bibinfo {year} {1999})},\ \Eprint
  {https://arxiv.org/abs/https://doi.org/10.1063/1.479408}
  {https://doi.org/10.1063/1.479408} \BibitemShut {NoStop}%
\bibitem [{\citenamefont {Ikeda}\ and\ \citenamefont
  {Tanimura}(2018)}]{IT18CP}%
  \BibitemOpen
  \bibfield  {author} {\bibinfo {author} {\bibfnamefont {T.}~\bibnamefont
  {Ikeda}}\ and\ \bibinfo {author} {\bibfnamefont {Y.}~\bibnamefont
  {Tanimura}},\ }\bibfield  {title} {\enquote {\bibinfo {title} {Phase-space
  wavepacket dynamics of internal conversion via conical intersection:
  Multi-state quantum \uppercase{F}okker-\uppercase{P}lanck equation
  approach},}\ }\href
  {https://doi.org/https://doi.org/10.1016/j.chemphys.2018.07.013} {\bibfield
  {journal} {\bibinfo  {journal} {Chemical Physics}\ }\textbf {\bibinfo
  {volume} {515}},\ \bibinfo {pages} {203--213} (\bibinfo {year}
  {2018})}\BibitemShut {NoStop}%
\end{thebibliography}%

\end{document}